\documentclass[iop,twocolumn]{emulateapj}

\newcommand{\Msol}{M$_{\odot}$}

\defcitealias{Kirk11}{Paper~I}
\defcitealias{Masch10}{M10}
\defcitealias{Krumholz10}{K10}

\begin{document}

\title{Variations in the Mass Functions of Clustered and Isolated Young
	Stellar Objects}
\author{Helen Kirk\altaffilmark{1} \& Philip C. Myers\altaffilmark{1}}
\altaffiltext{1}{Radio and Geoastronomy Division, Harvard Smithsonian
	Center for Astrophysics, MS-42, Cambridge, MA, 02138, USA;
	hkirk@cfa.harvard.edu}

\begin{abstract}
We analyze high quality, complete stellar catalogs for four young 
(roughly 1~Myr) and nearby (within $\sim$300~pc) star-forming regions:
Taurus, Lupus3, ChaI, and IC348, which have been previously shown
to have stellar groups whose properties are similar to
those of larger clusters such as the ONC.  
We find that stars at higher stellar surface densities
within a region or belonging to groups 
tend to have a relative excess of 
more massive stars, over a wide range of masses.
We find statistically significant evidence for
this result in Taurus and IC348 as well
as the ONC.  These differences correspond to having
typically a $\sim$ 10 - 20\% higher mean mass in the more clustered
environment.
Stars in ChaI show no evidence for a trend with either
surface density or grouped status, and there are
too few stars in Lupus3 to make any definitive interpretation.
Models of clustered star formation do not 
typically extend to sufficiently low masses or small group
sizes in order for their predictions to be tested
but our results suggest that this regime is important to consider.
\end{abstract}

\section{INTRODUCTION}
Does the distribution of masses of stars forming in isolation differ
from those forming in clustered environments?  The initial mass function
inferred from local field stars appears to be consistent with that
seen in clusters \citep[e.g.,][]{Bastian10}, however, the local
field star population is composed of a combination of stars which formed in
isolation, stars which dispersed from unbound clusters, and 
stars which were ejected from
bound clusters.  Most stars are believed to form in clustered
environments \citep[e.g.,][]{Lada06}, therefore stars which formed
in isolation or small groups may not be the dominant contributor
to the local field star population.

Differences in the distribution of masses of stars forming in clusters
versus isolation may be expected, particularly at higher masses.
Massive stars are known to form at least primarily within clusters, 
and it is uncertain
both observationally and theoretically whether massive stars can ever
form alone\footnote{The term ``massive'' typically refers to an OB
star, with a mass of several \Msol, however, in this paper, we also
consider slightly lower mass stars, with spectral types as late
as G.}.
In the competitive accretion scenario, the most massive stars 
start to form
early and spend most of their evolution in high density environments
in order to accrete sufficient mass to become a massive star;
lower mass stars later form in the gas around the massive stars 
\citep[e.g.][]{Bonnell08}.
Any massive stars found in isolation would therefore have moved there after 
accretion or have had their low-mass companions dispersed, 
rather than having formed in isolation.
Under the monolithic collapse scenario, stars form from 
the fragmentation of their natal core \citep[e.g.][]{McKee03}. 
Depending on the physical conditions, a massive core could fragment to
produce a massive star and / or a group of lower mass stars.  Massive
isolated stars are therefore not explicitly prohibited from forming by
the model, although they may be unlikely.  Cores
with column densities in excess of 1~g~cm$^{-2}$, where fragmentation
is suppressed \citep{McKee03}, have only been observed
in highly clustered environments.  In the stationary 
accretion model of \citet{Myers09}, massive stars accrete a significant
fraction of their mass from the `clump' material beyond their own
natal core.  To accrete sufficient mass within a sufficiently
short time, the clump gas must be denser than in isolated regions.
Low-mass
stars, on the other hand, could easily form in the surrounding
lower density, filamentary distributions of gas; it would 
be difficult to form a massive star in isolation.

Observationally, few, if any, massive stars appear
to have formed in isolation.  An upper limit of isolated
massive O stars in our galaxy was measured by \citet{deWit04,deWit05}
to be $4 \pm 2\%$, although this search was limited to massive
(O and B star) companions.  The apparently isolated O
stars may in fact belong to small clusters with lower mass
companions, as suggested by the models of \citet{Parker07}.

Clustered star formation appears to extend down to surprisingly small
size scales.  \citet[][hereafter Paper~I]{Kirk11} analyzed fourteen
small stellar
groupings, typically with 20-40 members, in four young, nearby
star-forming regions where deep spectroscopic catalogs are available. 
Despite possessing surface densities an order of magnitude
lower than the standard cluster, \citetalias{Kirk11} showed that 
these small stellar groups shared many of the
properties associated with clusters.  These properties include 
a correlation of the mass
of the maximum mass member with the total group mass, and the
central location of the most massive member.
The unique advantage of studying clustered star formation in these
small stellar groups is the lack of source confusion (due to a
combination of the lower source densities and closer distances
compared to typical clusters), the inclusion of very low
mass members (the catalogs are complete to late M spectral type),
and the young age (roughly 1~Myr).  Since only approximately
half of the stars in each region are identified as belonging
to stellar groups, while the other half of the stars are
found in relatively isolated environments, 
these data also provide an excellent
opportunity to study differences between the clustered
and isolated young stellar populations, in particular
the distribution of masses.

Therefore, in this paper we investigate the 
mass distributions of clustered and isolated populations of young
stars.  We address the question of whether more massive
stars preferentially form in clustered environments.
Our main conclusion is that higher mass stars are preferentially
found in clustered environments, over a wide range of stellar 
masses.  In order to minimize uncertainties introduced by estimating
stellar masses, much of our analysis is performed comparing
the spectral types of the stars, which have been
accurately measured.

In Section 2, we describe the stellar catalogs and the completeness
levels in each.  In Sections 3 and 4, we compare the distributions of 
spectral types for stars in high and low surface density environments
(Section 3) and those belonging to groups versus isolated stars
(Section 4).  We make comparisons to model predictions and various
observations in Section 5, discuss the results and interpretations
in Section 6, and conclude in Section 7.  Stellar motion is analyzed
in Appendix~\ref{app_pm}, and several additional statistical
tests are discussed in Appendix~\ref{app_mst}.

\section{DATA}
Currently, four nearby, young star-forming regions 
(Taurus, Lupus3, ChaI, and IC348) have excellent stellar 
catalogues with completeness to around 0.02~\Msol\ or type M8.5.
The Taurus data was originally compiled in \citet{Luhman10},
while the Lupus3 data is a combination of several source
lists given in \citet{Comeron08}.  In our analysis here, we exclude
the six stars which lie at a right ascension of less than 16h 05m and a
declination below 39\degr 50\arcmin, as we expect the stellar population
is incomplete here -- this area was covered by only one the source
lists in \citet{Comeron08} (his Table 8).  The ChaI data is
primarily from \citet{Luhman07}, and the IC348 data is primarily
a combination of the catalogs given in \citet{Lada06} and 
\citet{Muench07}.  In all cases, the spectral types of the
stars were determined spectroscopically, with a typical uncertainty
of roughly half a spectal subtype.  The total number of stars in each
region above the spectral completeness limit is 344 in Taurus, 58 in Lupus3,
215 in ChaI, and 349 in IC348.
These data are discussed in more detail in
Appendix A of \citetalias{Kirk11}.

In \citetalias{Kirk11}, we identified stellar groups based on
a minimal spanning tree (MST) analysis.  In a MST,
all stars are linked together by their nearest neighbours in a tree
diagram; groups are defined where all members are connected 
by separations less than the critical length, $L_{crit}$. 
Following \citet{Gutermuth09}, $L_{crit}$ was
determined based on the distribution of all branch lengths.
Appendix~D of \citetalias{Kirk11} examines the effect of
$L_{crit}$ on the groups identified, and shows that group
membership is little affected by variations of $L_{crit}$ within
the range of errors expected.
In \citetalias{Kirk11}, we set a minimum group size of 
more than ten members, 
based on a visual examination of the MST groupings.
Unlike the O-star studies of \citet{deWit04} and \citet{Lamb10},
foreground and background contamination of these catalogues is
not a concern, allowing groups to be identified without
relying on statistical comparisons with background source density
counts.

Masses were estimated based on a combination of stellar evolutionary
models by \citet{Palla99}, \citet{Baraffe98}, and \citet{Chabrier00},
and assuming a constant age of 1~Myr.  As discussed in \citetalias{Kirk11},
these assumptions lead to an overall uncertainty in the mass of each
star of about 50\%; the ranking of the masses is, however,
much more certain (and is exact for any single stellar age assumed).  
Where possible in the analysis here,
comparisons are made using spectral types rather than masses to
minimize the uncertainties.

\subsection{Orion Nebula Cluster}
In \citetalias{Kirk11}, we also compared the properties of
the young stellar groups to those of a typical young cluster.
For this purpose, we analyzed the ONC dataset of \citet{Hillenbrand97}.
There, our MST algorithm identified one very large cluster
(the central cluster where the Trapezium stars are found),
as well as five small stellar groups with properties
similar to those found in the four nearby star forming
regions.  In \citetalias{Kirk11}, we considered only the 
large ONC cluster for comparison, as it alone represented
typical cluster properties.

For the ONC cluster identification and analysis in \citetalias{Kirk11}, 
we included only
sources listed with a 70\% or higher `probability of membership'
(based on proper motion observations) in \citet{Hillenbrand97}.
We used this conservative cut to prevent potential contamination
by non-ONC members.  Many of the 
fainter sources in \citet{Hillenbrand97}, however, did
not have proper motion data available at the time, and hence
had no probability of membership measure.
\citet{Hillenbrand97} estimated that the majority of sources 
without proper motion data were likely bona fide cluster members, 
based on the fraction of known members as a function of
both the separation from the cluster centre
and spectral type.

A comparison of the properties of isolated and clustered
stars could be strongly sensitive to variable completeness, both
spatially and spectrally.  In order to avoid any potential bias in
our results for the ONC, we therefore run our analysis on two versions
of the ONC catalog.  The conservative proper motion-cut catalog 
analyzed in \citetalias{Kirk11} will be referred to as ONC-c;
additionally, we analyze the full ONC catalog with no cuts applied, which
will be denoted as ONC-f.  The former should have minimal contamination
from non-ONC stars, but an irregular completeness at later spectral types, 
while the latter catalog has better completeness but a higher likelihood 
of contamination.
We ran the same group-identification algorithm on both ONC catalogs
separately; due to the higher surface density, ONC-f has a smaller
$L_{crit}$ than ONC-c: $L_{crit}$(ONC-f) is 0.08~pc  and 
$L_{crit}$(ONC-c) is 0.13~pc.

\citet{Hillenbrand97} estimated that their optical catalog
contained roughly half the total number of cluster members
(the others being detected only in the infrared due to high
extinction), and that 
the optical sample was representative of the full distribution 
both in terms of spatial distribution and spectral types.  
Sixty percent of the optical sample
had spectral types measured, with a roughly uniform
completeness level both spatially and as a function of
I and V band photometry (assumed to roughly correlate with spectral
types).  While the spectral type determinations
are therefore not complete, \citet{Hillenbrand97} argue
that they should be {\it representative} of the full population
in every respect.
The ONC survey completeness is roughly 14.5~mag in K-band and 17.5~mag 
in I band in the least-sensitive part of the survey \citet{Hillenbrand97}, 
which corresponds to roughly a mass of 0.04 to 0.055~\Msol\ at an age
of 1~Myr or 0.06 to 0.075~\Msol\ at 5~Myr.  We adopt a conservative estimate
of a spectral completeness limit of M6.5. 

In general, we find similar results using either the ONC-c or ONC-f catalog,
suggesting the completeness is reasonably consistent in the ONC-c
catalog.

\section{DISTRIBUTION OF SPECTRAL TYPES IN HIGH VERSUS LOW SURFACE DENSITY
	ENVIRONMENTS}

\subsection{Calculating the Local Stellar Surface Density}
In order to compare the properties of stars inhabiting isolated 
versus clustered environments, a scheme to classify each is
required.
One simple method which does not rely on the definition of stellar
groups is to use the local stellar surface density, $\Sigma$.  
If the 
separation from the star to its $n$th nearest neighbour (where
$n$ is 1 for the star itself)
is $r_{n}$, then the local surface density is 
\begin{equation}
\Sigma = \frac{n-1}{\pi r_{n}^2} ~~~.
\end{equation} 
As discussed in \citet{Gutermuth09} and \citet{Casertano85},
the fractional uncertainty in $\Sigma$ varies as $(n-2)^{0.5}$; 
higher values of $n$ give a lower spatial resolution, but
smaller fractional uncertainty.  \citet{Bressert10} recently used 
this surface density measurement to argue that there is no
distinct scale for YSO clustering within nearby star-forming regions.

We calculate $\Sigma$ for
each star using both $n = 4$ and 9, to allow us to 
examine the dependence on $n$.  Since \citetalias{Kirk11}
found many examples of stellar groups with roughly ten to twenty
members, $n$ values larger than about 10 would have poor sensitivity
to this clustering, while $n$ values of two or three could  
bias surface density measures for close visual pairs or binary stars.

\subsection{Comparison of Spectral Type Distributions}
\begin{figure}[h]
\plotone{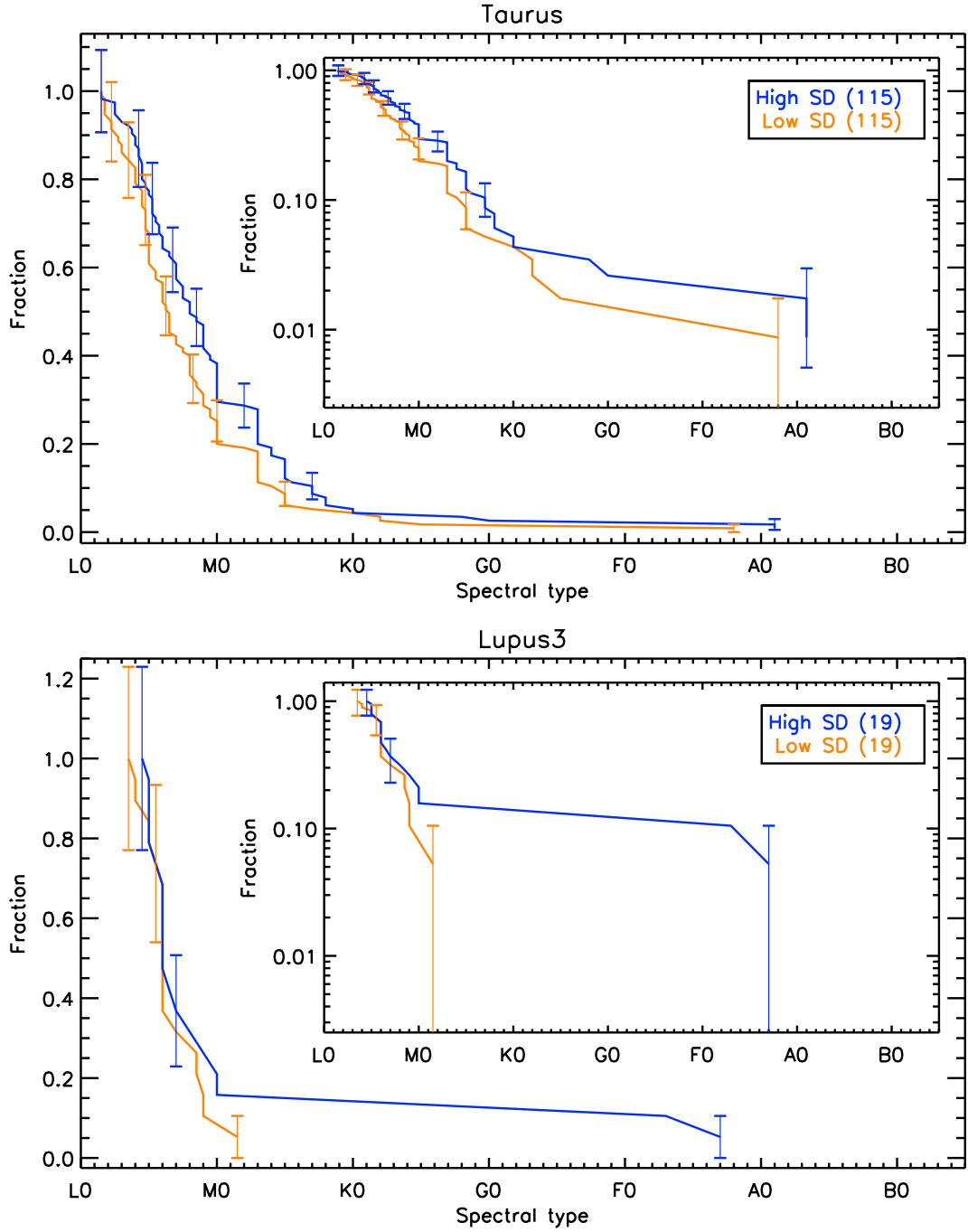}
\caption{The fraction of stars with spectral types earlier than each
	value for stars in the highest (blue) and lowest (orange)
	third of stellar surface densities, computed using $n=4$.
        The vertical lines indicate the Poisson error at each spectral type,
        and the numbers in braces in the legend indicate the total number
        of stars in each category.  The inset plot shows the cumulative number
        in log space.  Stars in Taurus and Lupus3 are shown in this
	figure. } 
\label{fig_cuml_sd1}
\end{figure}

\begin{figure}
\plotone{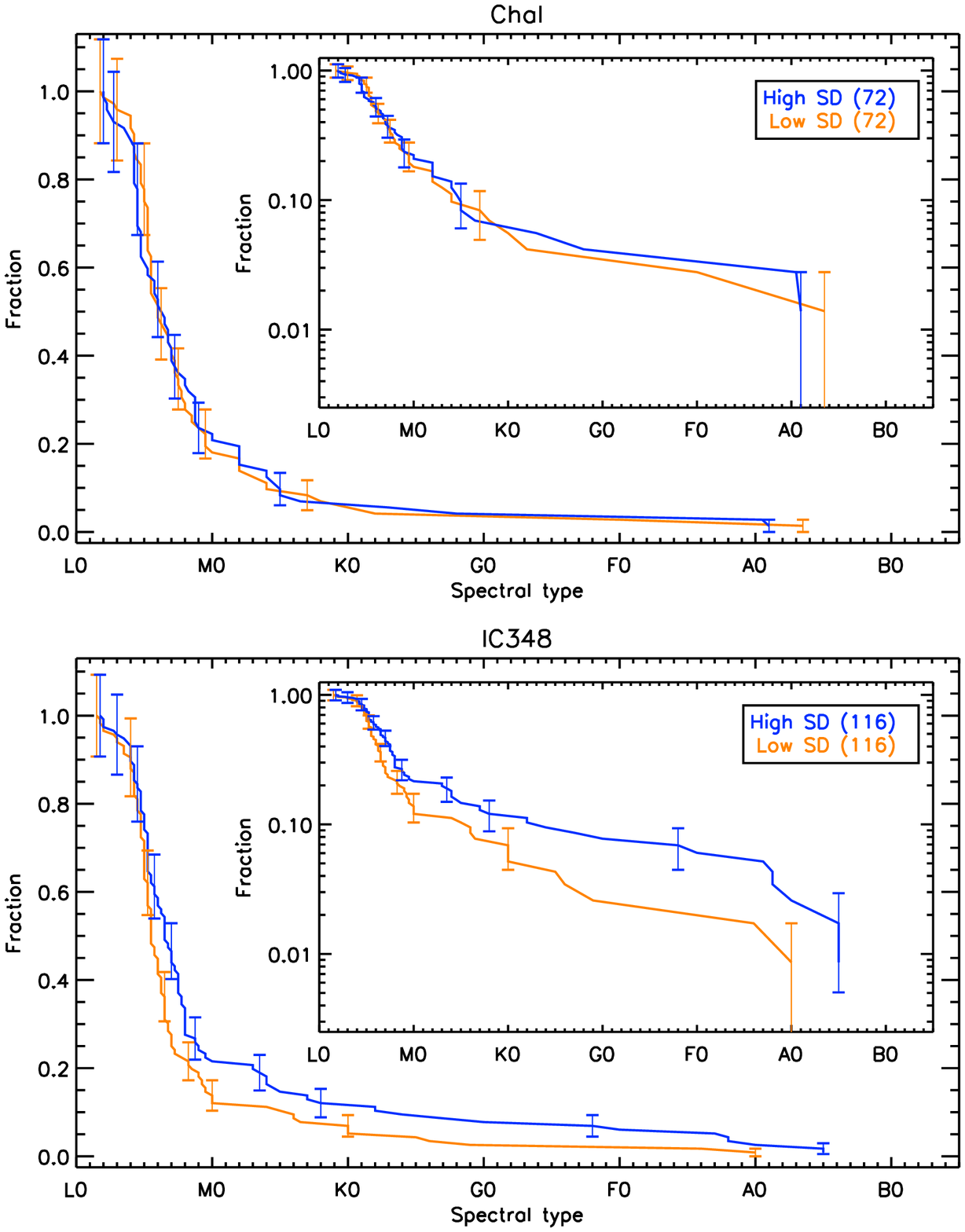}
\caption{The distribution of spectral types in the highest and lowest third
	of source surface densities in ChaI and IC348.  
	See Figure~\ref{fig_cuml_sd1} for more details.}
\label{fig_cuml_sd2}
\end{figure}

\begin{figure}
\plotone{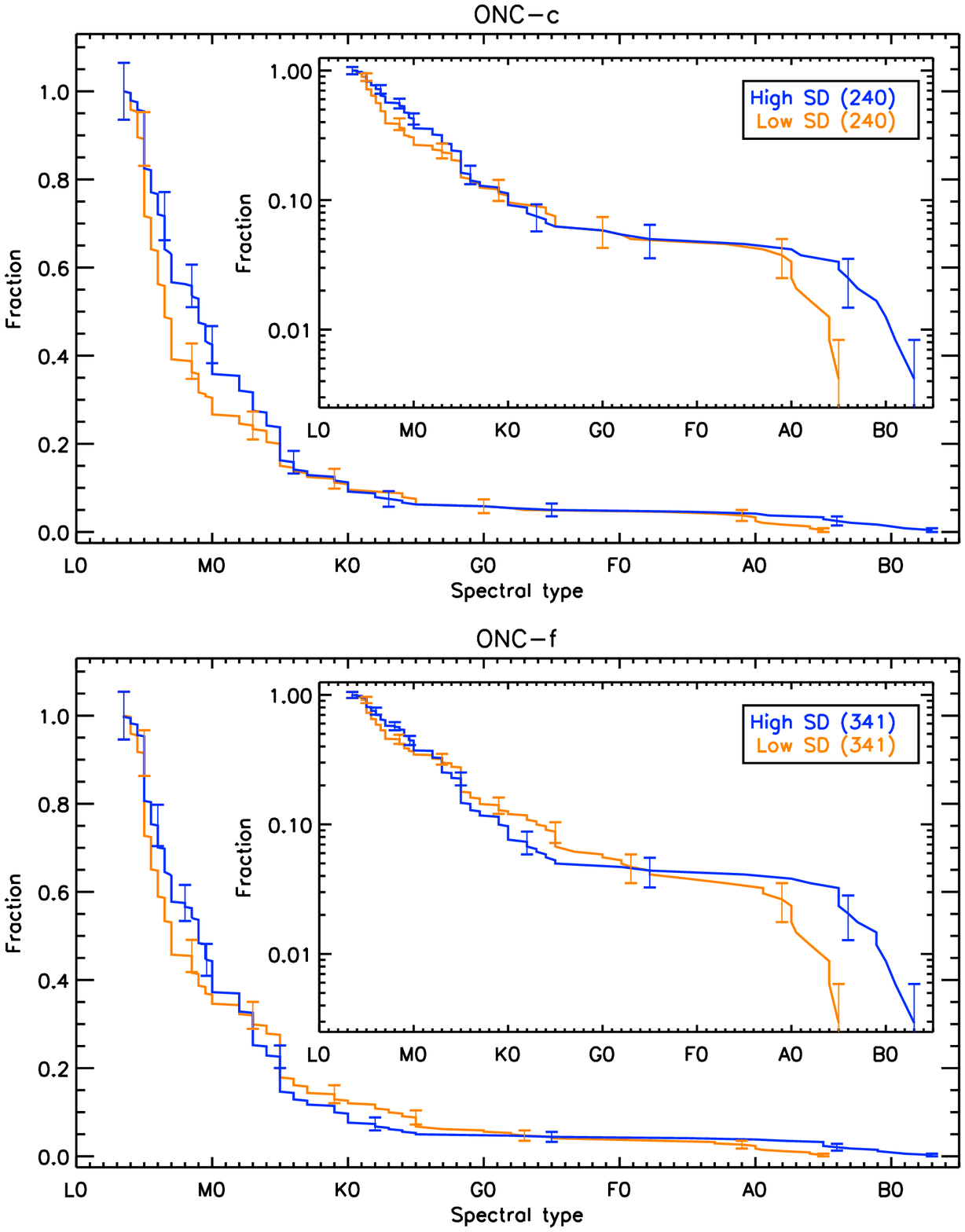}
\caption{The distribution of spectral types in the highest and lowest third
	of source surface densities in the ONC.
	See Figure~\ref{fig_cuml_sd1} for more details.}
\label{fig_cuml_sd3}
\end{figure}

We examine whether there is a global preference for earlier spectral 
types at higher $\Sigma$.  Ordering all YSOs within each region by
$\Sigma$ (for $n = 4$ and 9), we compare the spectral type
distributions for the upper and lower thirds of the population.
Figures~\ref{fig_cuml_sd1} to \ref{fig_cuml_sd3} show a 
comparison of the cumulative spectral type distributions for
the $n=4$ case for each of the four regions, as well as
the ONC for comparison.  Each plot shows the cumulative fraction in both
linear space (main plot) and log space (inset), to highlight differences
as the later and earlier spectral types respectively.  Comparisons are
only made for stars above the completeness levels discussed in
Section~2 and with known spectral types, as are all subsequent measures 
in this paper.  The error bars in the figures denote the Poisson
$\sqrt{N}$ error for each point.  Three of the regions -- Taurus, IC348,
and the ONC -- show striking differences between the high and low
surface density stars.  These regions show an overabundance of
stars at earlier spectral types in the high surface density environment
versus the low surface density
environment.  This is not only the case at the earliest spectral
types, but extends throughout nearly the entire range of spectral
types in the sample.  Statistical tests described below
confirm this visual impression.  The results of the statistical
tests are given in Table~1. 
Similar trends, but with poorer statistical
significance were found comparing the upper and lower $\Sigma$ halves
of the population, likely because stars at intermediate values of
$\Sigma$ have similar properties, but are distributed
in both bins.

\subsection{Statistics}
We ran three statistical tests on the distributions of spectral
types for the grouped and isolated stars to quantify
the differences visually suggested in Figures~\ref{fig_cuml_sd1}
to \ref{fig_cuml_sd3}.  The first two of these tests examine
global differences between the stars in high and low surface
density environments,
which will therefore be weighted where the bulk of the stars are,
i.e., the later spectral types.  The final test
focusses on the differences solely at early
spectral types.

\subsubsection{Kolmogorov-Smirnov test}
We first ran a two-sample Kolmogorov-Smirnov (KS2) test;
the KS2 test provides an effective way to measure whether
two datasets are statistically similar, i.e., whether it is
likely that both are drawn from the same distribution.  The
KS2 test is sensitive to differences in the datasets'
medians and variances \citep{Conover99}.
Table~1 shows the KS2 probabilities that the
high and low surface density stars share a common distribution
of spectral types.  The KS2 test suggests that the high and
low surface density stars 
are distinct at the 95\% or higher confidence
level with either $n = 4$ or 9 in IC348 and in the ONC, 
while in Lupus3, the populations are 
consistent with being drawn from the same distribution
at the 96\% confidence level, likely
due to the small number of stars in that region.  
ChaI has intermediate KS2 probabilties and so
does not fall definitively into one or the other category.
Taurus shows a strong difference in populations for $n = 4$
(95\% confidence level), but not for $n = 9$, suggesting
that very small stellar groupings are significant in the spatial
distribution of stars in Taurus.

We ran similar comparisons between stars in high and low
surface density environments with more severe (earlier) completeness
levels assumed, and find similar KS2 values for completeness levels
of 3-5 spectral subtypes higher in nearly every region, for both the
$n = 4$ and
9 surface density measures.  The one exception is the ONC-c catalog,
where KS2 values remained low only for a completeness level of up to 
two spectral subtypes earlier than what we assumed.

The KS2 test does not have a formal mechanism to include
measurement errors in the calculation.  In order to assess the
effect of uncertainty in the spectral types on the KS2 test
statistic, we therefore ran a series of trials with random errors
added to the measured spectral types.  For each entry in 
Table~1, we ran 10,000 trials with added spectral type 
uncertainties of half a subtype, and calculated the resulting
KS2 test statistic.  This is summarized in 
Table~2, and demonstrates that the significant
KS2 test statistics are robust to uncertainty at the expected
level.  Systematic biases in the spectral types measured are
not expected as a function of spatial clustering or spectral
type (K. Luhman, private communication).

\subsubsection{Mann-Whitney}
While the KS2 test is an excellent way to determine whether
two distributions are different, it does not provide a measure for
{\it how} they differ, e.g., if one of the distributions tends to
have extra high- or low- valued members.  To address this, we ran a
second statistical test, the Mann-Whitney (MW) test,
also known as the Wilcoxon test \citep{Conover99}.  In this test, two
datasets are compared based on their relative ranking (i.e., the
ordering of the values of both datasets combined); tied values
are each assigned the average rank of the ties.
The basic premise of the test is that two sets of data drawn
from the same overall distribution will have roughly equal total
ranks, whereas if one data set tends to have higher values, then
the total of its ranks will tend to be lower (since a rank of 1
denotes the largest value).  
The MW test probability that the high surface density stars tend to have 
earlier spectral types than the low surface density stars
is given in Table~1 for
each region.  {\it We emphasize that the MW test is most sensitive to
differences in the range of spectral types where the bulk of the
stellar populations are, i.e., earlier spectral types refers
primarily to those at the late-type end, around M-type.}
Taurus, IC348, and the ONC show high probabilities that the stars
in high surface density environments tend to have earlier spectral types.
Both Lupus3 and ChaI have no strong evidence that either the
higher or lower surface density stars tend to be of earlier spectral
type.

Where differences are seen in the spectral type distributions,
we can make a rough estimate of this typical difference.  We apply a 
global shift to the spectral types of the low surface density stars,
and measure how large an offset may be added so that the MW test
shows negligible statstical significance that the high surface 
density stars tend to have earlier spectral types.  We find that
required shifts in the spectral types are quite small - shifting all of
the low surface density stars earlier by between half and one spectral
subtype is sufficient for the MW test to return probabilities of 
85\%\footnote{The value of 85\% is chosen to represent a probability that,
while reasonably high, is low enough to prevent any firm conclusions.}
or smaller that Taurus, IC348, and ONC stars have earlier spectal types in 
high surface density environments.

As with the KS2 test, we ran the same statistical comparisons varying
the assumed completeness level of the sample, and found the same result,
i.e., the conclusions drawn from the MW test remain unchanged for
completeness levels up to at least two spectral sub-types earlier
than assumed.

Uncertainties in the spectral types measured also do not have
an impact on the MW test statistic -- we ran a similar test to that
described in the previous section for the KS2 test, and find
little variation in the MW test statistic (see
Table~2).

\subsubsection{Early Type Stars Counting Test}
Neither of the above tests explicitly focusses on differences
specifically in the early type star populations at high and low
surface densities.  There are an insufficient number of stars in each region
(save the ONC) to run the above two tests only on the
early type stars in each environment.
Instead, we must rely on a simpler statistical test, comparing the total
number of early type stars in high and low surface density environments.  
If the populations were identical, roughly the same number of
early type stars should be found in the high and low surface density
samples, since the total number of stars in each is equal.  To
determine whether the differences are significant, we compare the
value to the Poisson error in the number of early type stars in the
low surface density sample.
We then define the excess of early type stars in the high surface
density environment as 
\begin{equation}
\begin{array}{rcl}
E_{early} & = & \frac{N_{early,high~sd} - N_{early,low~sd}}
{\sqrt{(\sqrt{N_{early,high~sd}})^2 + (\sqrt{N_{early,low~sd}})^2 }}  \\ 
 & = & 
\frac{N_{early,high~sd} - N_{early,low~sd}}{\sqrt{2 N_{early,low~sd}}} \\
\end{array}
\end{equation}
which gives roughly the number of sigma of significance.

The value of $E_{early}$ depends on what spectral type is used as the
early type cutoff.  Since it is not obvious, a priori, what spectral
type should be used, we performed the calculation for a range of cutoffs,
every half a spectral type from late G to early B, in order to see what
type(s) would give the highest statistical significance to the difference.
Table~1 gives the maximum value of $E_{early}$ we found,
and the early type cutoff spectral type or types that it was found for.
The four nearby regions tend to have $E_{early}$ values of $< 2 \sigma$
with an early type cutoff of around G0.  The ONC had much higher
$E_{early}$ values, from 3 to 8 $\sigma$, when a very early type cutoff,
around B5, was used.

\subsubsection{Statistics Summary}
In summary, we find that there are statistically significant
differences between the spectral types of stars found in
low and high surface density environments in IC348 and the ONC,
as well as Taurus, when small stellar groupings are
considered (using $n = 4$).  In these regions, the higher
surface density stars have an overabundance of earlier spectral
types compared to the lower surface density stars.
This overabundance corresponds to an average spectral
type which is half to one spectral subtype earlier for the
high surface density stars.
There are too few stars in Lupus3 to show any statistically
significant differences, while the ChaI stars do not
follow the general trend.
Focussing only on the early type stars, none of the four nearby
regions have a sufficient number of stars to show statistically
significant differences in the populations.  Significant
differences are seen, howevever, in the ONC, as has also been
noted in previous studies \citep[e.g.,][]{Hillenbrand98}.

\subsection{Combined Distribution}
\begin{figure}
\begin{tabular}{c}
\includegraphics[height=3.5in]{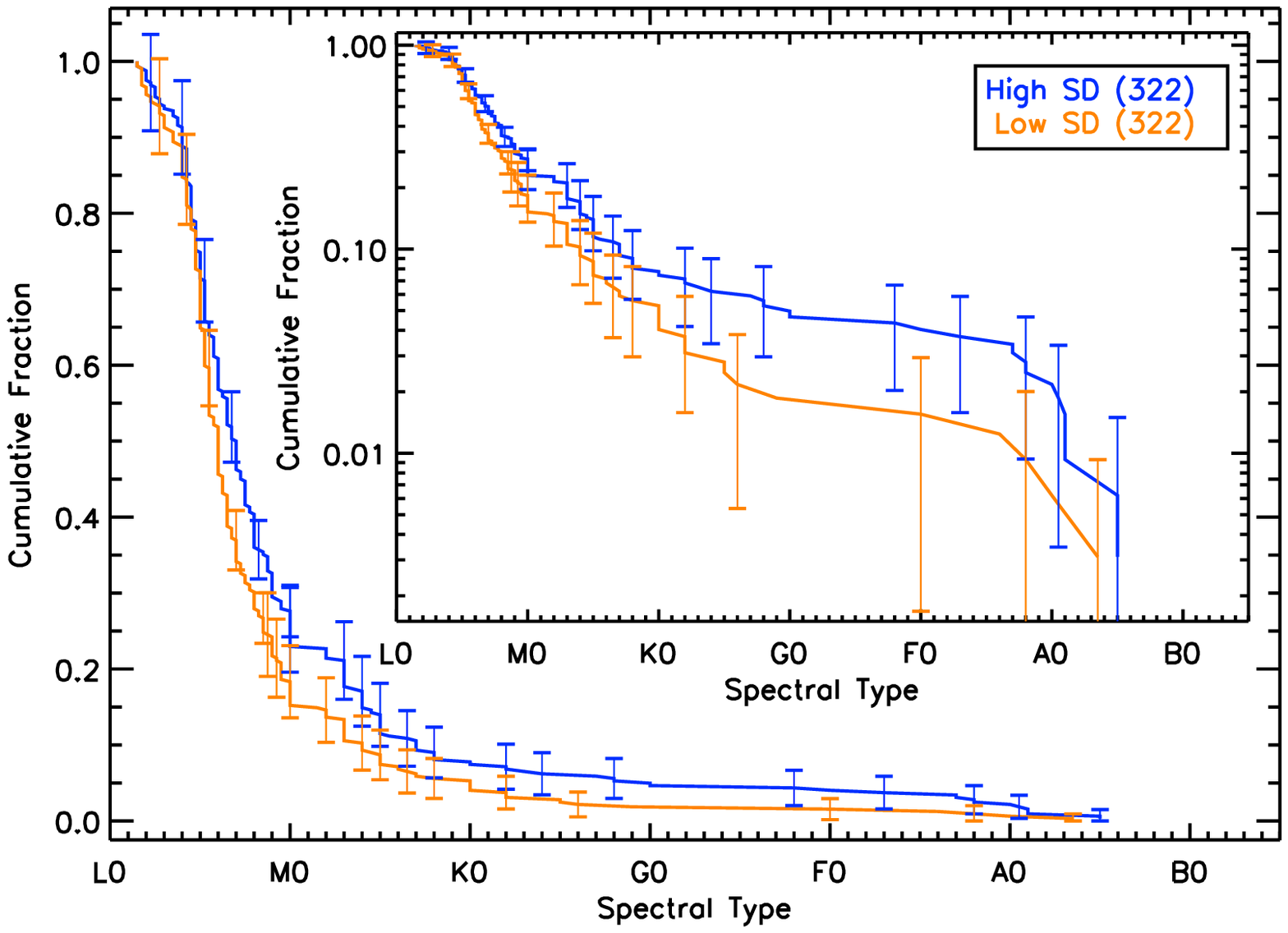} \\
\includegraphics[height=3.5in]{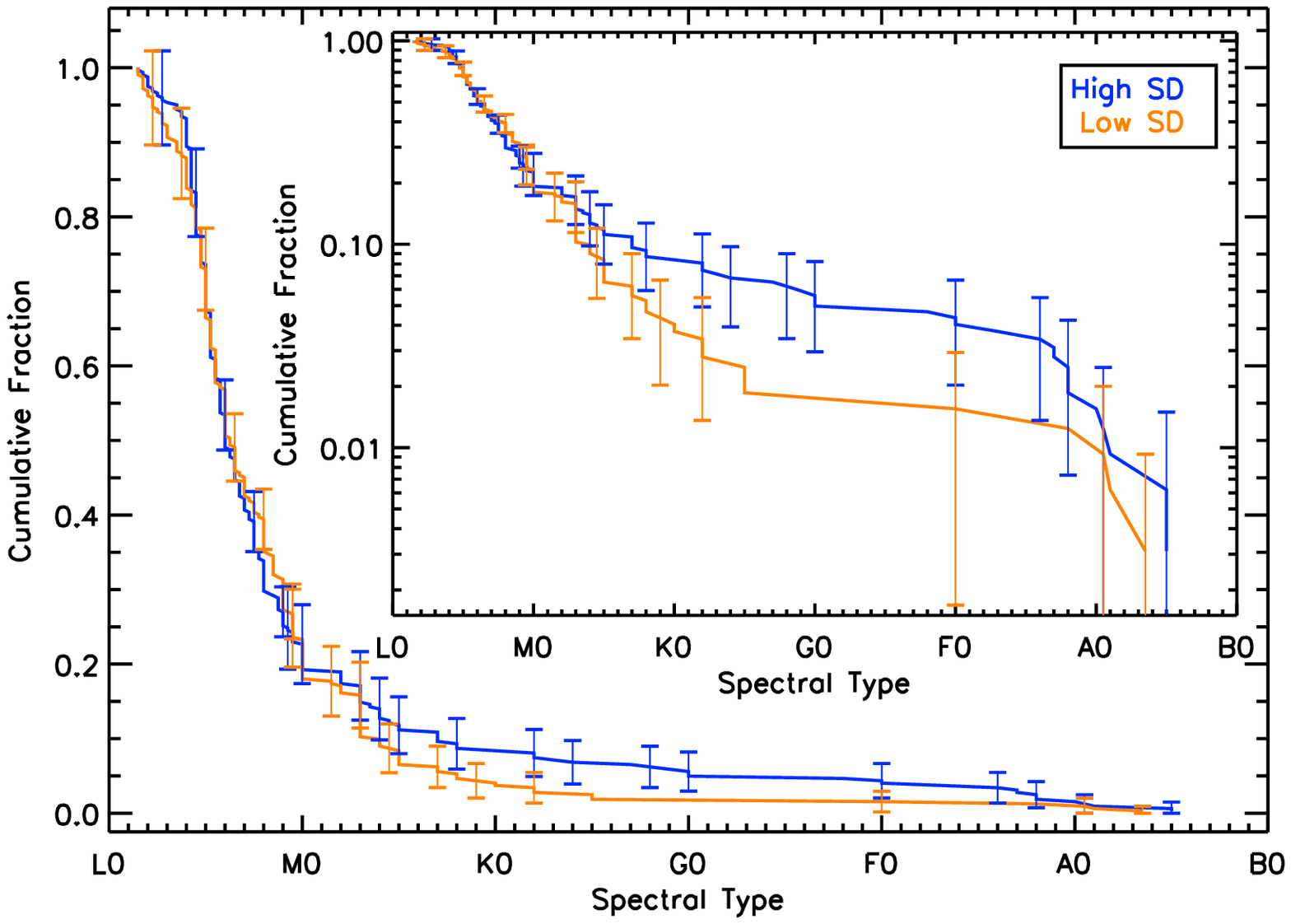} \\
\end{tabular}
\caption{The distribution of spectral types in the highest and lowest
	third of surface densities combined over Taurus, Lupus3, ChaI, and
	IC348.  The top panel shows the distributions when combined
	using the divisions in Figures~\ref{fig_cuml_sd1} and 
	\ref{fig_cuml_sd2}.  The bottom panel shows the distributions
	when divided using the highest and lowest third of the
	surface densities compared across all four regions.}
\label{fig_cuml_sd_all}
\end{figure}

Finally, we examine the spectral type distributions for the combination
of all four regions.  Combining the spectral type distributions may
increase the statistical significance of the difference between the high
and low surface density environments, since all regions but ChaI visually 
follow the same trends.  Merging the individual spectral type distributions
can be done in two ways.  The first method is to add together the spectral
types for all of the YSOs classified as high surface density in their
individual regions, and similarly for the low surface density stars.
This is shown in the top panel of Figure~\ref{fig_cuml_sd_all}, and the 
resulting statistical measures
are given in Table~1.  The appearance is qualitatively
similar when stars in ChaI are excluded, however, the statistical
significance of the trends is slightly higher in this instance
(these numbers are also given in Table~1).

On the other hand, the data from the four regions can also be combined
by splitting up the stars into those falling in the upper and lower
{\it global} thirds of the surface densities measured in all four
regions.  This would be the more appropriate combination method
to adopt if the mechanism driving the difference between 
environments was dependent on the absolute
surface density of stars.  The bottom panel of Figure~\ref{fig_cuml_sd_all} 
shows the combined
spectral type distributions split by absolute surface density.
Note that each region populates a different fraction of the 
high and low surface density cuts; Taurus YSOs dominate the low surface 
density cut, while IC348 YSOs dominate the high surface density cut.
The difference between the spectral type distributions using the
two combination methods is striking.  While both show an excess
of early type stars in the higher surface density environment,
the size of the difference and the range of spectral types over
which the difference is present are both smaller when the global
surface density cut was applied.  The smaller range of the excess
has a significant effect on the KS2 and MW statistics measured,
since much of the population is at later spectral types where the
difference between high and low surface densities is diminished.

The fact that combining the data from the four regions using the relative
surface density cuts rather than a global cut shows a much
larger difference between the two populations suggests that the 
mechanism responsible for the difference at high and low surface 
densities operates in a relative sense, rather than coming into
effect at a set absolute surface density value.  It is interesting to note
that the global surface density comparison shows a much stronger
difference when {\it Taurus} rather than ChaI stars are excluded
from the sample, despite Taurus stars on their own showing a difference
while ChaI stars do not.  This appears to be 
the case because Taurus's early type stars, which are in relatively
higher surface density environments within Taurus, are classified
as being in low surface density environments in comparison to
surface densities seen in the other regions.

\section{DISTRIBUTION OF SPECTRAL TYPES IN GROUPS VERSUS ISOLATION}
\label{sec_grp_isol}
\begin{figure}[htp]
\plotone{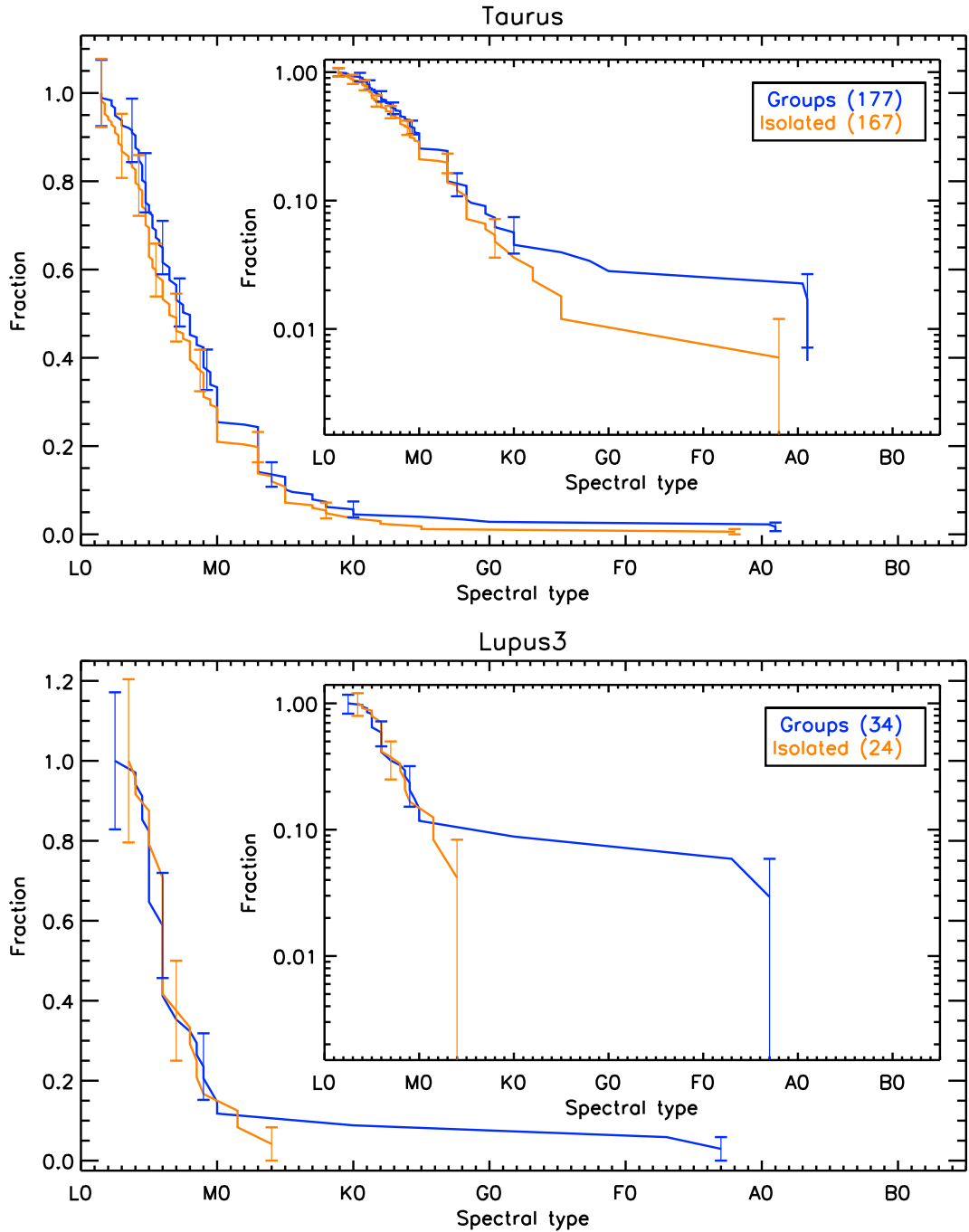}
\caption{The distribution of spectral types for grouped and isolated
	stars.  See Figure~\ref{fig_cuml_sd1} for the plotting
	conventions used.
	Stars in Taurus (top) and Lupus3 (bottom) are shown in this figure.}
\label{fig_cuml_spec1}
\end{figure}

\begin{figure}
\plotone{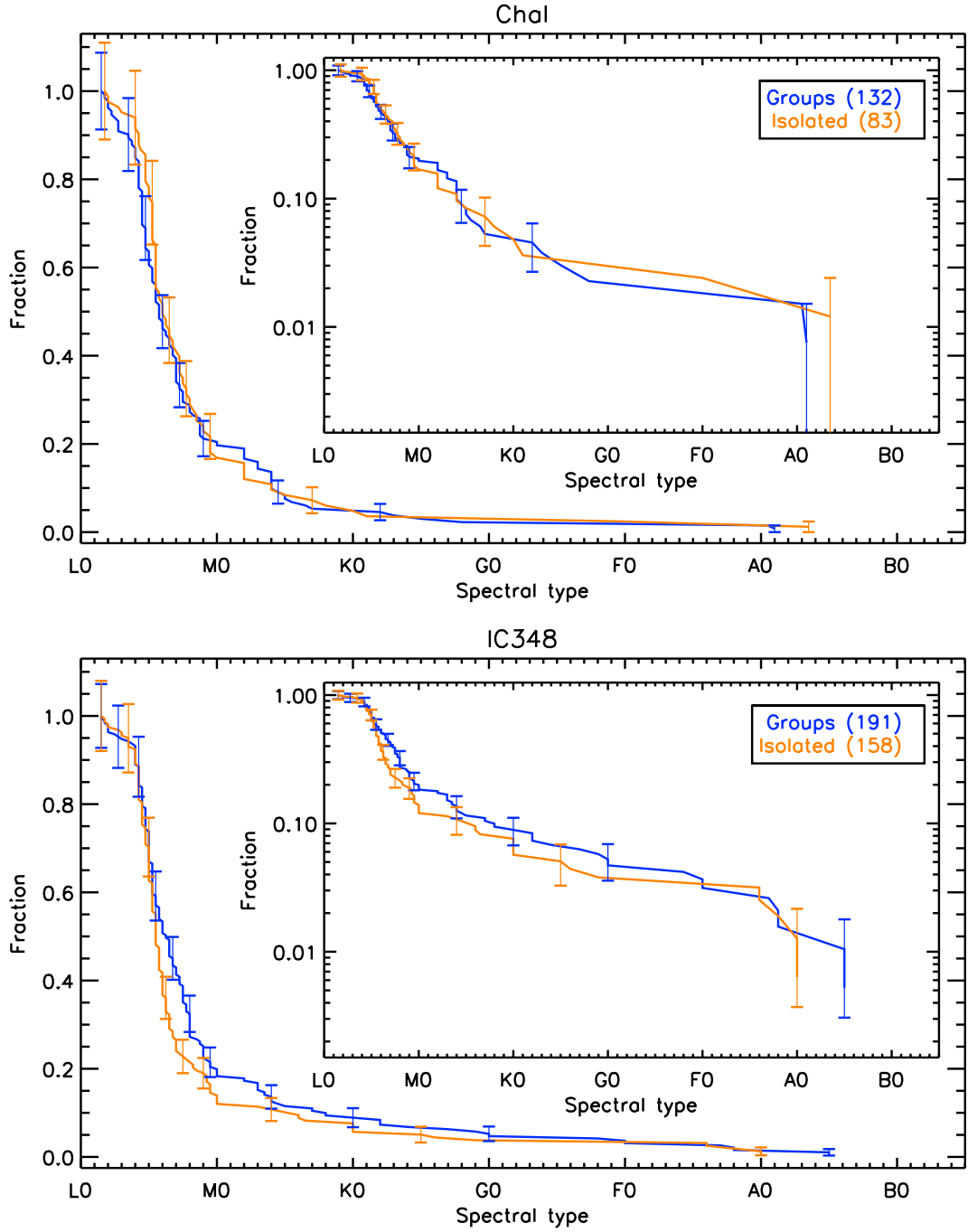}
\caption{The distribution of spectral types for grouped and isolated stars.
	See Figure~\ref{fig_cuml_sd1} for the plotting conventions.  
	Stars in ChaI (top) and IC348 (bottom) are shown in this figure.}
\label{fig_cuml_spec2}
\end{figure}

\begin{figure}[htp]
\plotone{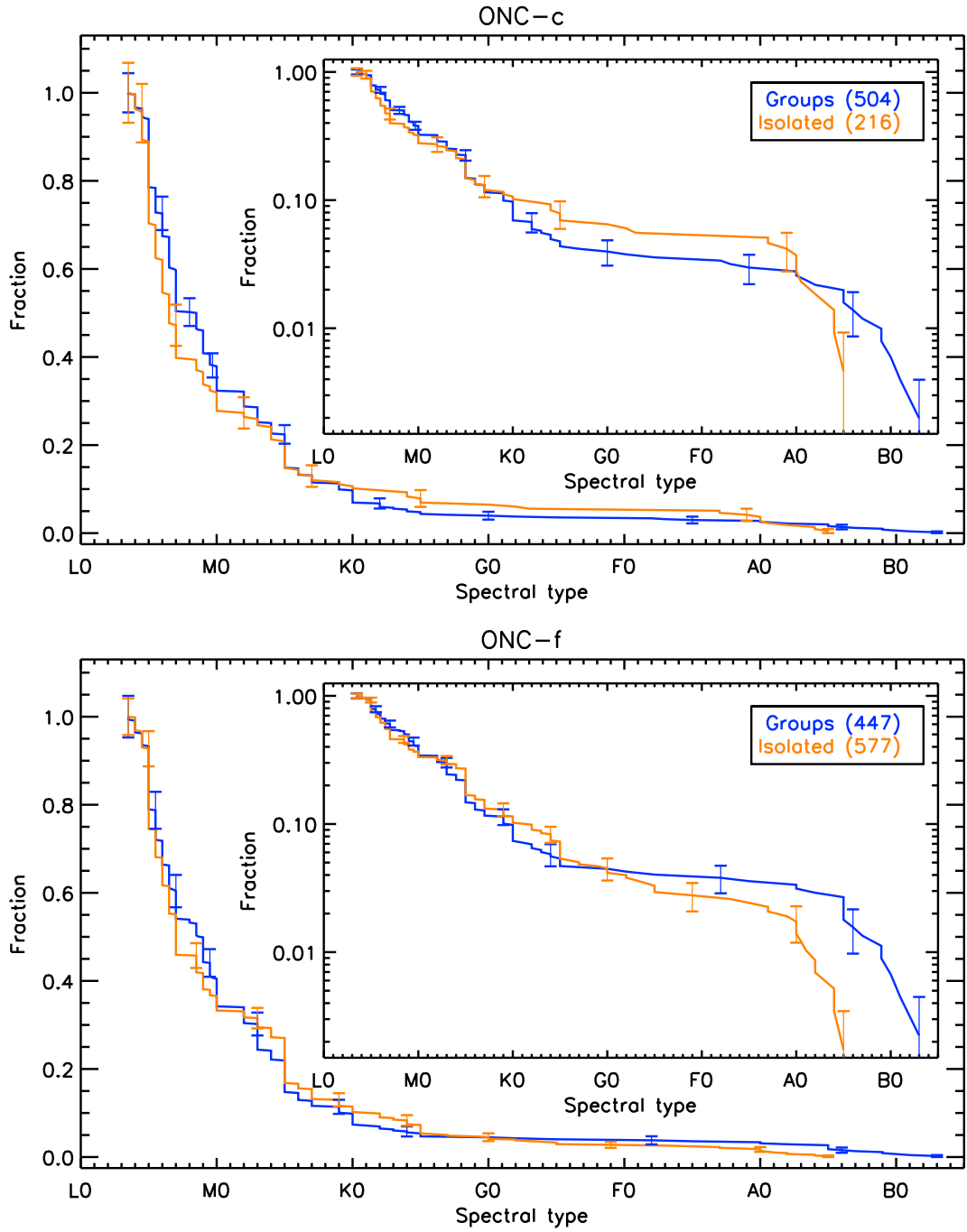}
\caption{The distribution of spectral types for grouped and isolated stars.
	See Figure~\ref{fig_cuml_sd1} for the plotting conventions.  
	Stars in the ONC are shown in this figure.}
\label{fig_cuml_spec3}
\end{figure}

A second way to compare the stars is to subdivide the stars in
each region into those that were found to belong to an
MST-defined stellar group in \citetalias{Kirk11} and those
that were not associated with a group.  As discussed in Section~2,
groups are defined as having more than ten members with nearest
neighbour separations less than $L_{crit}$; stars not associated
with a group (`isolated stars') are therefore all stars which
are connected to fewer than ten sources with separations of
$L_{crit}$ or less.
Figures~\ref{fig_cuml_spec1} to \ref{fig_cuml_spec3} show the 
cumulative fraction of sources earlier than a given spectral
type for each of the four regions as well as the ONC for comparison,
using the same plotting conventions as Figure~\ref{fig_cuml_sd1}.  

Unlike the comparisons at high and low surface densities,
the grouped versus isolated cumulative distributions tend to show
significant differences over more localized spectral types.
IC348 in particular, and Taurus to a lesser extent, appear to show 
an excess of later-type stars (mid-M) in the groups, and this also
appears in the ONC.
The main difference between the
two ONC catalogs is the stars in groups with spectral types
around F.
\citet{Hillenbrand97} found that stars around
type F were the most likely to have small proper motion
membership probabilities; if proper motions were preferentially
measured for stars nearer the centre of the ONC, then our
ONC-c catalog would have a deficit of F stars in groups versus
isolation compared to the full ONC-f catalog, as 
Figure~\ref{fig_cuml_spec3} shows. 

\subsection{Statistics}
We ran the same statistical comparisons as discussed in Section~3
on the grouped and isolated stellar distributions.  These results
are summarized in Table~3.

In general, we see similar results to the statistical
tests for the high versus low surface density stellar distributions,
although with generally slightly poorer statistical significance.
We again verified that the spectral completeness limits adopted do not
affect the results, for completeness levels at least 2-3 subtypes
earlier than we assumed.

Here, to compare the number of the early type stars in groups verus
isolation, we first correct for the total number of stars in
each category.  The number
of early type stars expected to be found in groups,
based on the number of early type stars found in isolation, is
\begin{equation}
N_{early,grp,exp} = N_{early,isol} \times \frac{N_{tot,grp}}{N_{tot,isol}}
\end{equation}
and the excess of early type stars in groups over that expected from
the isolated population is analogous to equation 2, but using the full
error propogation required from equation 3.

Similar to the high and low surface density populations, we
find relatively small excesses of early type stars.  The excesses
are typically smaller than for the high and low surface density
populations, since the extra step of normalizing the total sample 
sizes increases the associated error measure.  Here, the largest 
magnitude of excess
in ChaI is actually a {\it deficit} of early type stars in groups,
although at less than 1$\sigma$ significance.
The ONC stars again show the most significant excesses of early
type stars in the groups, but only at the $\le 2 \sigma$ level.
We tested the effect of measurement uncertainty in the
spectral types using the same procedure as outlined
in Section~{3.3.1}.  The results, listed in Table~4,
show that the KS2 and MW test statistics are robust to the
spectral type error.

In summary, similar to the surface density comparisons, the 
tests show that IC348 and the ONC have statistically significant
differences between the spectral types of stars in groups versus
isolation.  In Taurus, IC348, and the ONC, stars in groups have a 
{\it global} tendency to be at earlier spectral types than stars
in isolation, again with a high statistical significance.  
This tendency is not obvious when examining only the earliest 
spectral types (with the $N_{early}$ test), likely due to small
number statistics.  The trends seen in these three regions
tend to be slightly
statistically weaker than those seen comparing the high and low surface 
density environments, and may be more pronounced at localized
spectral types.  Lupus3 again does not have a sufficient
number of stars to see a statistically significant difference,
and ChaI does not appear to follow the general trend.  

\subsection{Combined Distributions}
\begin{figure}
\plotone{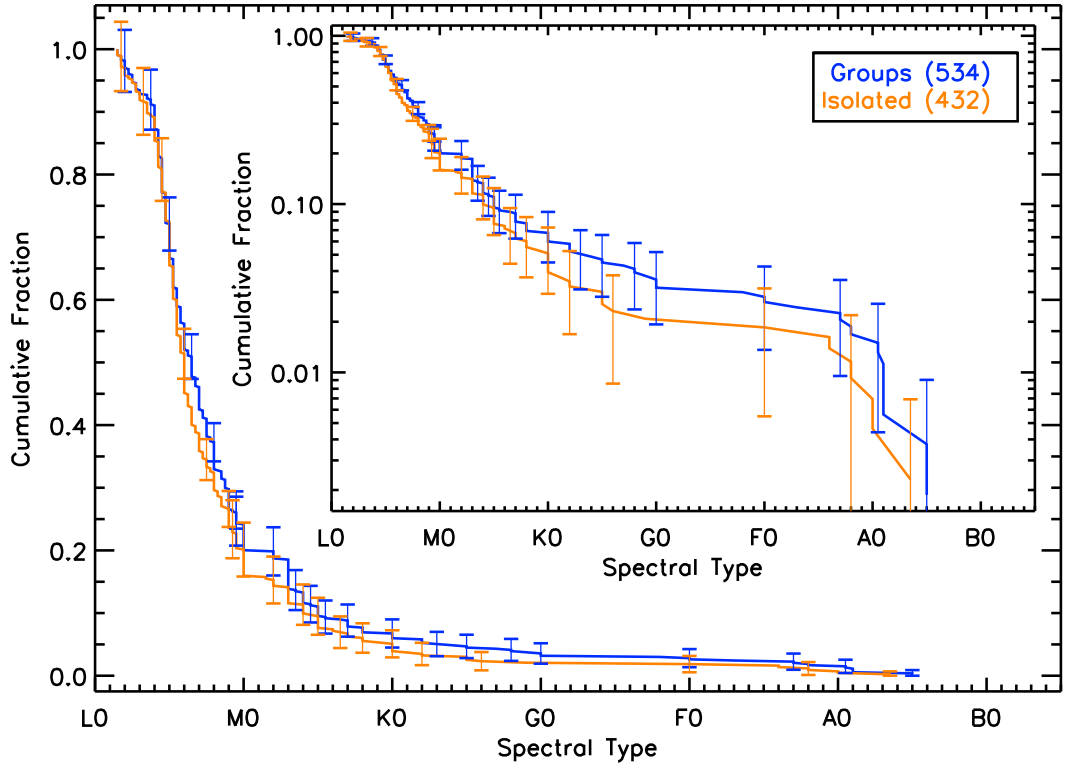}
\caption{The distribution of spectral types for grouped and isolated
	stars summed over all of Taurus, Lupus3, ChaI, and IC348.}
\label{fig_cuml_grp_all}
\end{figure}

Similar to Section~3.4, we also compare the distribution of stellar
spectral types for stars in and outside of MST groups in 
Taurus, Lupus3, ChaI, and IC348 combined.  This data is shown in 
Figure~\ref{fig_cuml_grp_all}, and the corresponding KS2 and
MW statistics are given in Table~3.  As expected,
combining the data from the four regions tends to enhance the
trends seen, as is also apparent in the statistical measures.
Excluding the stars in ChaI, where no trend is apparent,
has little effect on the qualitative appearance of
Figure~\ref{fig_cuml_grp_all}, but does slightly further strengthen
the statistical measures.

\subsection{Individual Taurus and ChaI Groups}

While the populations of grouped stars are dominated by a single
clustered environment in both IC348 and the ONC, and only one
group was identified in Lupus3, in the Taurus grouped population, there
are comparable contributions from eight different stellar groups.
Here, we examine each group independently to determine whether our
conclusions change when groups are treated separately.
Similarly, we
compare the three groups identified in ChaI, noting that there,
one of the groups dominates the total number of sources by a factor
of two.

The number of stars in each group above the spectral completeness
limit ranges from fourteen to thirty one in Taurus, and from twelve
to eighty three in ChaI, which are small sample sizes to search for 
statistically
significant differences.  Nevertheless, we compared the KS2 and MW
test statistics for each group individually versus all of the isolated
stars in the region.  These statistics are given in 
Table~5.  Results from the $N_{early}$ test
are not shown, as none of the groups have a statistically
significant excess of early type stars {\it at the earliest
spectral types}, due to the small number of sources
in each group.  

\begin{figure}[h]
\plotone{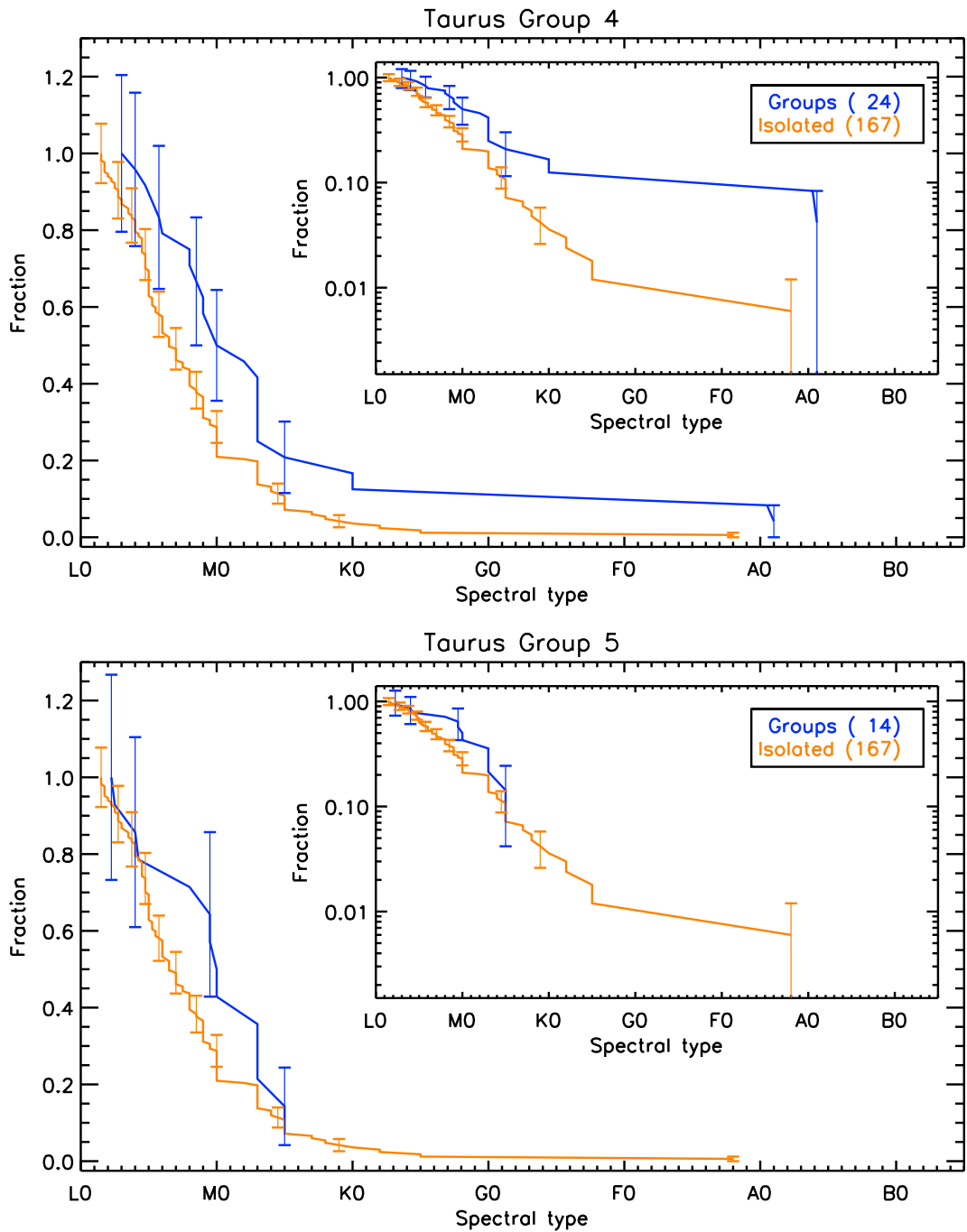}
\caption{The distribution of spectral
	types in the two Taurus groups showing the most significant
	differences with the isolated populations (based on KS2 and
	MW statistics).  The plot conventions follow those in
	Figure~\ref{fig_cuml_sd1}.}
\label{fig_Taur_cuml_spec}
\end{figure}

Most of the Taurus groups have KS2
probabilities indicating the single group and isolated stars are 
statistically indistinguishable, and MW probabilities that neither
population has a strong tendency for more massive members.  Surprisingly,
two Taurus groups did show strong {\it global} differences with 
the isolated
population, despite the small number of members.  Both Taurus Groups 4 and 5
(L1551 and L1529) have KS2 probabilities of less than 10\%
and MW probabilities of $\gtrsim90\%$, implying that overall
the stars in these groups tend to have earlier spectral types than
the isolated population.  
The spectral type distributions of the exceptional Taurus
Groups 4 and 5 are
shown in Figure~\ref{fig_Taur_cuml_spec}.

Group 5 / L1529 is a small group, with its most massive member
only of spectral type K5, and shows a substantial excess of sources 
relative to the isolated population around type M0.  Group 4 / L1551,
on the other hand, is a somewhat larger group, with a B9 / B9.5
pair of stars, and shows a systematic offset of more massive stars
at all spectral types. 
In \citetalias{Kirk11}, L1551 was one of only two groups which
did not have a centrally-located most massive member -- the
B9 and B9.5 stars fall near the outskirts of the group.
L1551 therefore is only an exemplar for the differences seen
in the spectral type distributions in groups and isolation,
and not also the location of the most massive member.  In 
\citetalias{Kirk11}, we noted that had the group been defined
with a slightly smaller $L_{crit}$, the B9/B9.5 stars would
not have been considered group members, and
the next most massive member {\it is} centrally-
located.  Interestingly enough, if the B9/B9.5 stars were excluded
from the L1551 spectral type distribution, the grouped and
isolated stars would still show a systematic difference -- the
absence of the two B9/B9.5 stars is not sufficient to make
the two distributions agree. 

Conversely, we test whether the excess of earlier type
stars is confined to only these two groups in Taurus.
We compare the spectral type distributions for the stars
in all groups except groups 4 and 5 and the isolated stars, and
find a diminished, but still visible, excess of early type
stars in groups.  This excess only extends down to about type
K0, and is sufficiently small that the KS2 and MW tests do
not yield significant results.  This appears to indicate that
the bulk of the difference in the grouped stars populations
arises from only a few of the groups, although there are hints
that the difference is present at a lower level in at least
some of the other groups.

In ChaI, each individual group appears much more similar to the
isolated distribution; none of the groups have the striking
visual appearance of differences that are seen in 
Figure~\ref{fig_Taur_cuml_spec}.  ChaI Group 2 has a relatively
low KS2 probability that the grouped and isolated stars are
similar (12\%); this appears to be caused by a slight {\it deficit}
of grouped stars at mid-M spectral types versus the isolated
distribution.

\subsection{Distribution of Spectral Types with Varying Group Size}

Our analysis in Section 4.1 distinguishes groups
and isolated stars assuming that groups must have more than
ten members.  The MST group identification scheme used in 
\citetalias{Kirk11}, however, identifies all groupings
of stars which are separated by less than $L_{crit}$,
and therefore identifies groupings as small as
pairs of stars.  The minimum group size is somewhat
arbitrary, chosen in \citetalias{Kirk11} to allow for the measurement
of mass segregation.
In light of the fact that very small groupings
of stars may be important, particularly in Taurus (Section~3), it is useful
to examine various minimum group sizes.
A visual examination of the YSOs suggests that a significant
fraction of the earliest type stars not found in our $N > 10$ MST groups
are instead found within smaller MST groupings; there are relatively
few early type stars which do not belong to any size of MST grouping.

We re-ran the statistical comparisons discussed in Section~4.1
on grouped and isolated stars using different minimum group
sizes.  The details are discussed in
Appendix~\ref{app_mst}; the overall conclusion
is that the definition of a group
does not have a strong influence on the 
differences seen between the grouped and isolated stellar
spectral types.  Increasing the minimum
group size substantially causes some of the
genuine groups to be classified as isolated sources,
diluting the difference between the two populations.
Decreasing the minimum group size has
a small effect on the statistical measures
until the minimum group size is
so small that few stars are categorized as isolated and
the ability to statistically distinguish the two 
populations is poor.
Table~6 shows the results for the same
statistical tests as given in Table~3 but
with a minimum group size of $N \geq 2$ and $N \geq 15$.
As in Section~3, the stars in Taurus show the
largest increase in the KS2 probability with an
increase of the minimum group size, suggesting that
stellar clustering on small scales is important in this region.

\section{DISTRIBUTION OF MASSES}
\label{sec_models}

Comparison between our observations and other observations,
in addition to predictions from 
clustered star formation models are more easily made using
the distributions of mass, rather than spectral types.
As illustrated in Sections 3.4 and 4.2, the distribution
of spectral types combined over the four nearby star-forming 
regions offers a reasonable representation of the trends seen
in the individual regions, with a generally higher
statistical significance due to the greater number of
stars compared.  For simplicity in mass space, we make 
comparisons only of these combined datasets.
Note that for the surface density comparison,
we use the combination displayed in the top panel of 
Figure~\ref{fig_cuml_sd_all}, i.e., each region contributes an equal
number of YSOs to the high and low surface density populations.

As discussed in \citetalias{Kirk11}, our simple method 
of assuming a constant stellar age tends to introduce
biases in the mass distribution.  To better understand
this effect, we made comparisons of the mass distributions 
shown in \citet{Luhman09}
for Taurus, ChaI, and IC348 versus the mass distribution
which we would derive using the same basis set of spectral
types and a constant 1~Myr age.  
We find that the assumption of a global 1~Myr age
tends to cause an overestimation of the number of the most
massive stars in each region (around 2~\Msol\ and above),
and hence an accompanying underestimate of the number of slighly less
massive stars;
there is also a smaller prevalent overestimate
of the number of stars with around 0.2~\Msol, and a similar
underestimate at slightly lower masses.  
Assuming a constant age of 2~Myr instead,
we find significantly improves our mass estimations above
$\sim$1~\Msol; we could not find a single age estimate to
improve the smaller bias at the lower mass end.  
For our analysis of the mass distributions we therefore
revise our masses above 1~\Msol\ to be those predicted 
assuming an age of 2~Myr, which slightly reduces the values.
We stress that our mass distributions should still be
viewed as approximate estimations, given our simple
age assumption.

\begin{figure}
\includegraphics[height=6.7in]{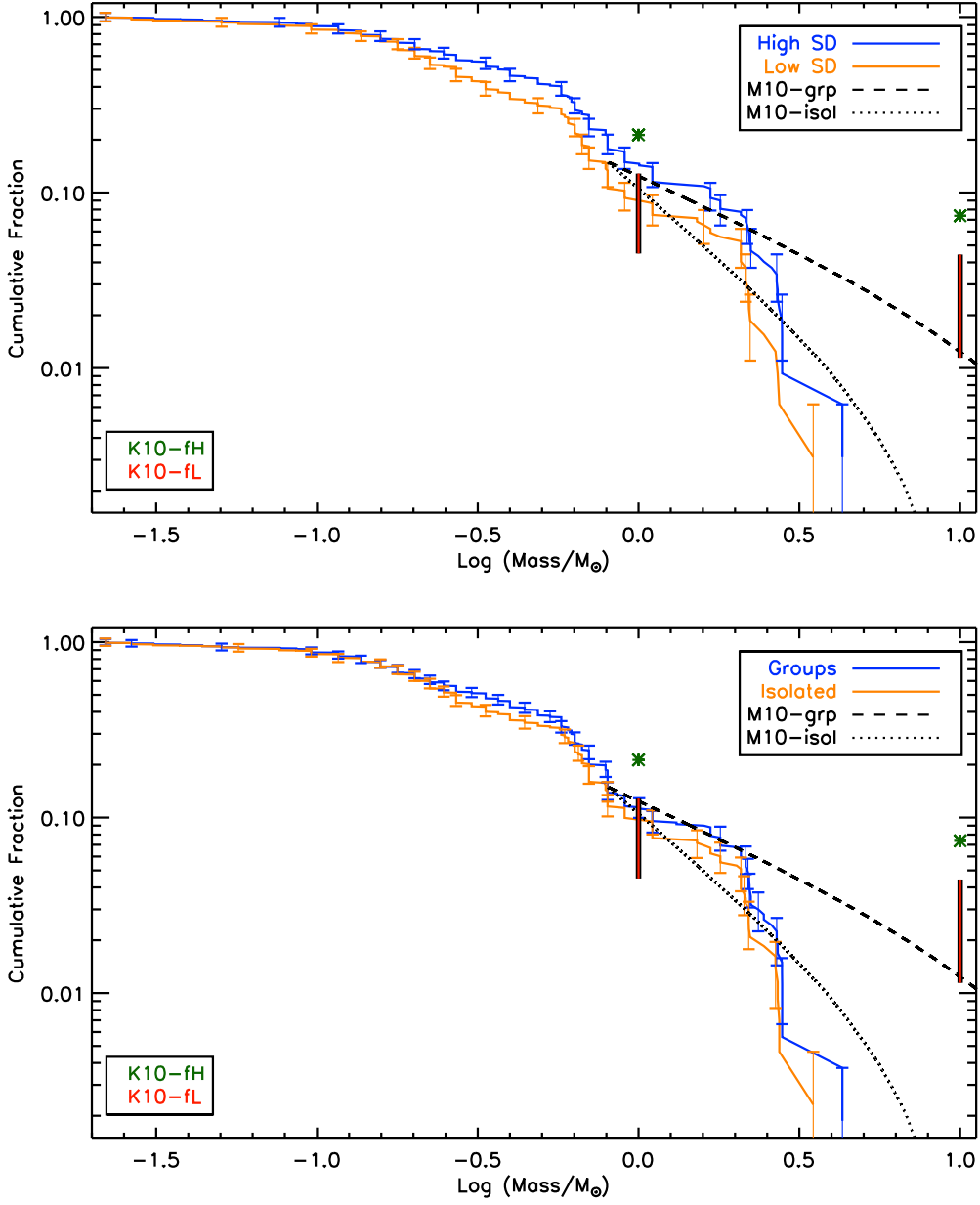}
\caption{The cumulative distribution of mass for all four nearby star-forming
	regions combined.  The top panel shows the high and low stellar
	surface density stars, while the bottom panel shows the
	grouped and isolated stars.  Error bars
	indicate the Poisson errors.  Also shown are
	predictions from \citetalias{Masch10} for stars in grouped
	and isolated environments, scaled to a total fraction of 15\%
	at the lower mass limit of 0.8\Msol (black dashed and dotted
	lines respectively).  The range of values spanned by the
	analytic \citetalias{Krumholz10} model are shown by the
	green asterisk (high density model) and red line
	(range of low density models).
	The \citetalias{Krumholz10} models are not directly comparable
	to the observations, however, the range spanned by the models
	{\it should} encompass the observations.}
\label{fig_combo_cuml_mass}
\end{figure}

Figure~\ref{fig_combo_cuml_mass} shows the cumulative mass
distribution for the regions, overlaid with predictions
from several models (more details below).  The differences
between the populations are less visually apparent than
in Figures~\ref{fig_cuml_sd_all} and \ref{fig_cuml_grp_all}; 
the difference is entirely due to
the variable stretch in the conversion from spectral to mass
space, since a single conversion between spectral type and
mass is assumed.  
The bumps in the mass distribution are likely due to 
inaccuracies in our assumed conversion.
Note that despite
the visual appearance, our statistical tests provide the same
results in mass-space as spectral type-space, since we
assume a one-to-one translation between the two.
To quantify the typical mass difference
between the two distributions, we ran a similar MW shift test to that
described in Section 3.3.2, finding the shift which, when applied
to the low surface density or isolated stars, returns an MW probability 
of $\sim$85\%.  We found that for both Taurus and IC348,
the high surface density stars are typically 18\% more massive
than the stars in low surface density environments, while stars 
in groups are 4\% and 11\% more massive than their isolated counterparts
in Taurus and IC348 respectively.

\begin{figure}[htp]
\plotone{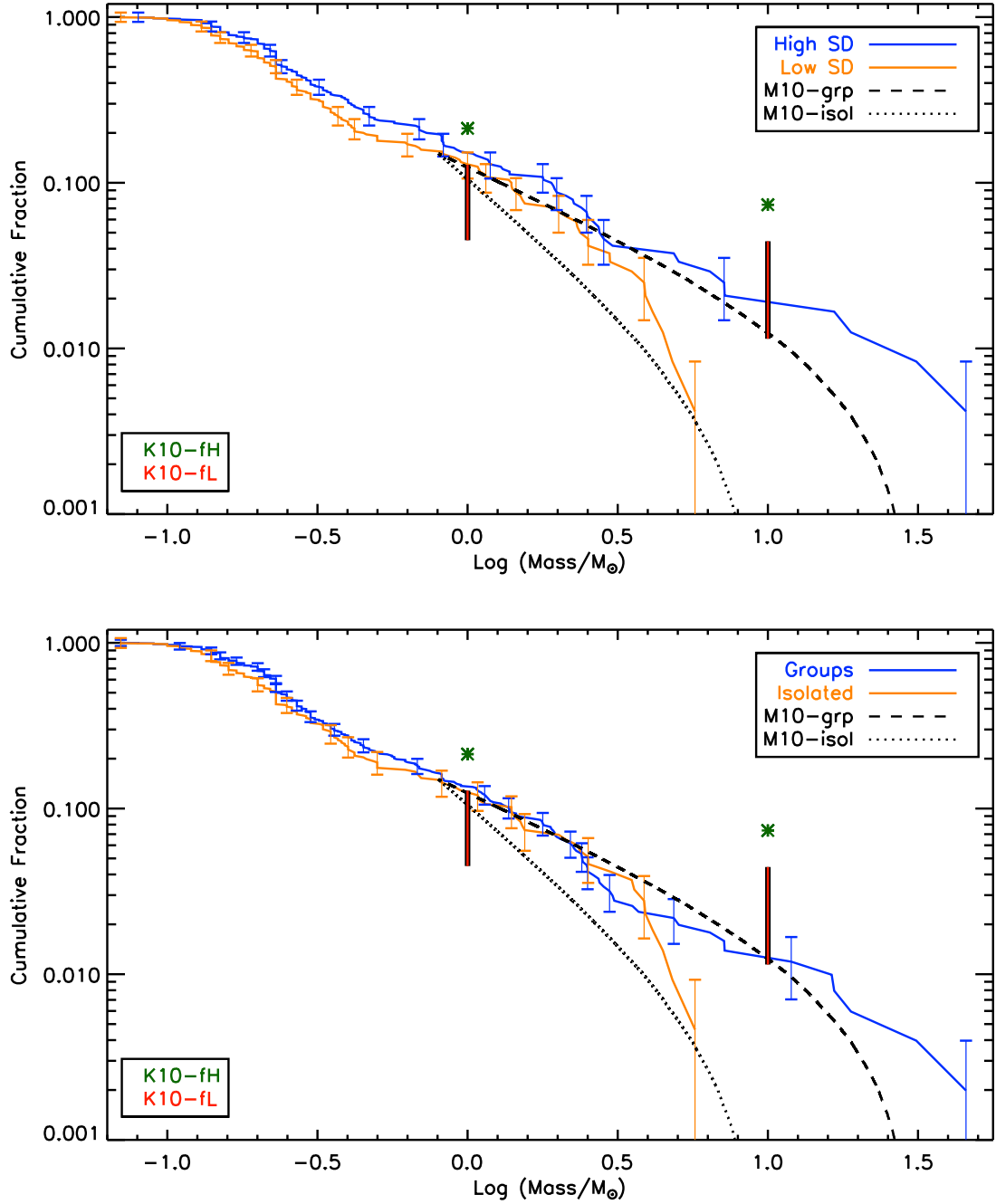}
\caption{The cumulative mass distribution for the stars in ONC-c, for
	the high versus low stellar surface density sources (top panel),
	and grouped versus isolated sources (bottom panel).  
	Comparisons with predictions by \citetalias{Masch10} and
	\citetalias{Krumholz10} are shown as in 
	Figure~\ref{fig_combo_cuml_mass}; note that the 
	plot ranges differ from the previous figure.}
\label{fig_cuml_mass_ONC}
\end{figure}

For comparison, Figure~\ref{fig_cuml_mass_ONC} shows the 
corresponding plots for the ONC-c catalog; the ONC-f catalog
appears similar.  In the ONC, the mass estimates are expected
to be more accurate, since \citet{Hillenbrand97} allowed the
stellar ages to vary when estimating the masses.
Applying the same MW shift test as above, we find that the stars
in high surface density environments are roughly 8\% more massive
than the stars in low surface density environments, while the
stars in groups are 6\% more massive than their isolated counterparts.

\subsection{Effect of Single Age on Statistics}
The statistics discussed above were calculated assuming a constant
stellar age, however, variations in the stellar ages could
affect the values quoted.  Here, we discuss additional tests to investigate
this effect.

\subsubsection{KS2 and MW Test Statistics}
As noted above, with the assumption of (any) constant stellar age,
the KS2 and MW test statistics are identical to those found for
the spectral type distributions, since there is a one-to-one
mapping between the two.  Allowing for age variations could
change the shape of the mass distributions.  In particular,
having systematically older stars the more clustered environments
would tend to decrease their masses, and therefore lessen
the difference seen between the more and less clustered
populations (i.e., high versus low surface density or grouped
versus isolated).  We ran tests varying the stellar ages adopted for
the clustered stars to determine the age difference size
necessary to make the two populations appear similar.  
Assuming that the less clustered population
is 2~Myr old,
then we find that the more and less clustered populations can match
at the very high mass tail (stars above $\sim$2~\Msol)
only when the more clustered stars are $\sim$5~Myr old.
Even this large an age difference, however, is
insufficient to decrease the excess of more massive stars in the
more clustered population in the $\sim$0.2 - 2~\Msol.
This remains true even if the less clustered population is
reduced to an age of 1~Myr.
Since stellar evolution tracks run roughly vertically for
masses below of order 1~\Msol, changes in age have almost
no impact on the stellar masses inferred from the spectral
type.

Since age thus primarily affects only the highest mass stars
which are few in number, the KS2 and MW statistical tests are
largely impervious to systematic changes in age.  For any
reasonable assumption about ages, the KS2
test statistic remains small (5\% or less) in all cases where 
it was small under the assumption of a global constant age.  Age
has a slightly higher, but still small, effect on the MW statistic.
In Taurus, the MW statistic drops by less than 2\% for more
clustered stars which are 1~Myr older than the less clustered
stars; only when the more clustered stars are set to 5~Myr while the 
less clustered stars are 1 or 2~Myr does the MW statistic drop 
below 90\% in some cases.  Even in this drastic case, the MW
statistic for the $n=4$ surface density comparison only changes
from 99\% to 98\%.
In IC348, the effect is slightly larger, due to a greater
number of sources at the higher mass ranges which are effected by
the age assumed.  Still, the MW test statistic drops by at most
a few percent when the more clustered stars are $\leq$ 1~Myr older
than the less clustered stars. 
All of the IC348 MW test statistics remain above,
and normally well above, 90\% while the more clustered stars are
younger than 5~Myr.

\subsubsection{Mass Shift Values}
We find that the mass shift values quoted are only very weakly
dependent on the stellar ages assumed.
We ran similar comparisons between the more and less clustered 
stars, and found that the mass shifts were roughly identical
for any constant age adopted for both popuations.
If the more and less clustered populations have ages different by
less than 0.5 Myr, again the mass shifts measured remain 
virtually unchanged.  For more clustered stars which are
older (/younger) by 1~Myr, the mass shifts are decreased (/increased)
by at most 2\%.  

\subsubsection{Summary}
The statistical results presented for the mass distributions are 
largely insensitive to variations in the age assumed, despite age being a 
strong factor of uncertainty in the absolute values of the masses
estimated.
In order to decrease the statistical significance of the results
presented in this section, stars would need to be systematically
and noticeably older (a $>$ 1 Myr difference) than stars in the less clustered
environments.
There is no clear physical mechanism which would cause more
clustered stars to be systematically older by such an amount.
In fact, clustering is often seen to
be tighter for younger stars than older ones \citep[e.g.,][]{Luhman10}.
Systematically younger stars in clusters would tend to strengthen
our statistical measures and would tend to increase the mass shift
measured between the more and less clustered environments.

\subsection{Comparison to Model Predictions}
Here, we compare the observed cumulative distribution
of masses to two recent model predictions
 -- the first, analysis of competetive accretion
simulations in \citet[][hereafter M10]{Masch10}, and the
second, analytic predictions by \citet[][hereafter K10]{Krumholz10} 
based on the monolithic collapse scenario.

\subsubsection{Competitive Accretion Simulations}
\citetalias{Masch10}
analyzed the clustering properties of stars (represented by sink particles)
formed in the competitive accretion simulations of \citet{Bonnell08}
and \citet{Bonnell03}, with a focus 
on the evolution of clusters and their most massive members.
Clusters were identified using the MST formalism,
with $L_{crit}$ set to 0.025~pc through visual inspection 
and a minimum size of thirteen members.
For comparison, in \citetalias{Kirk11}, we determined $L_{crit}$
based on the distribution of branch lengths, with $L_{crit}$
ranging from 0.52~pc in the more dispersed Taurus stars 
to 0.083~pc in the more tightly clustered IC348.  The \citetalias{Masch10}
critical length for highly clustered simulations therefore seems
reasonably consistent with our definitions; the minimum number of group
members is also similar.

\citetalias{Masch10} found that about 60\% of all sources form within
clusters, and of those sources, most tended to form with the
same radial distribution as already existed in the cluster.
The most massive sources (defined as any source with a 
final mass greater than about 1~\Msol), on the other hand, tended not to form
in clusters -- rather, clusters tended to later form
around a pre-existing massive source.  Often, this started
out as a small `sub-cluster', which then successively
merged with other small `sub-clusters', and eventually formed
a massive cluster, with the massive stars from each sub-cluster
ending up in the centre of the final system.  This evolution 
occurred rapidly, as the simulations were only run for 0.5~Myr.

At the end of the \citet{Bonnell08} simulation, the distribution of masses
of sink particles found within groups / clusters and in isolation
and the fits listed in Table~1 of \citetalias{Masch10}.
The sink particle masses are upper limits to the
masses expected in the stars -- the authors noted that effects 
such as stellar feedback and binary fragmentation were not
included in the simulation, and would serve to decrease the resultant
stellar mass.  With that caveat, Figure~\ref{fig_combo_cuml_mass}
shows the sink particle mass distributions in \citetalias{Masch10}
overlaid on the observed stellar distributions.
\citetalias{Masch10} only measure the mass distributions down
to 0.8~\Msol, which corresponds to roughly 15\% of the total
number of sources in all of our observations; the lines
plotted show the \citetalias{Masch10} results scaled down
by this factor.

A comparison between the scaled-down \citetalias{Masch10} distributions 
and the observed ones show that the two are in reasonable 
agreement, given both the
small number statistics (and estimated mass uncertainty) in
the observations, as well as the errors \citetalias{Masch10}
estimate in their mass function parameters (not plotted).  
It is also interesting to note that the \citetalias{Masch10}
results show a substantial distinction between the clustered
and isolated sinks, with most of the massive sinks being
found in clusters.  We observe a much smaller difference between
the stars in groups and isolation.
It is possible, however, that some processes which can reduce
the final stellar mass from that in the simulated sink particles
(such as stellar feedback) operate more strongly in a
clustered environment, which would decrease the difference
between the two \citetalias{Masch10} curves.

\subsubsection{Monolithic Collapse}
An analysis similar to \citetalias{Masch10} has not
yet been performed for the monolithic collapse scenario,
making a direct comparison to our observations challenging.
The best available predictions are from \citetalias{Krumholz10},
who construct several simple analytic models to describe the
distribution of final stellar masses based on results from
several star formation simulations which include radiative
feedback.

Based on both the simulations and a prior theoretical
prediction \citep{Krumholz08}, \citetalias{Krumholz10} 
find that in high density environments (above $\sim1$~g~cm$^{-2}$),
fragmentation of massive dense cores is largely inhibited,
allowing for the formation  of massive stars.
In lower density environments, fragmentation
occurs much more regularly, leading to smaller
stellar masses arising from a given core.
These results were used as the basis for an analytic 
model in \citetalias{Krumholz10} which predicts the final
distribution of stellar masses formed in both high and
low density environments.  \citetalias{Krumholz10}
assume the initial dense core mass distribution
follows the \citet{Chabrier05} IMF, scaled up by 
a factor of three, and a star formation efficiency
for any core fragment of a factor of 1/3.
In the high density case (model `fH'), 
they assume no fragmentation, i.e., each core forms one star,
and the mass distribution mirrors the \citet{Chabrier05}
IMF.  
In the low density regime (`fL'),
several different cases of fragmentation are
considered, with varying amounts of mass in each fragment.

The high and low density regimes of the \citetalias{Krumholz10}
models do not have a direct corresponendence to our observations,
as the spatial distribution of high and low density cores was
not specified in the model.  While a direct comparison therefore
cannot be made, the observed distribution of masses for grouped 
and isolated stars should fall within the range spanned by the models.
Figure~\ref{fig_combo_cuml_mass} shows the range spanned by the
\citetalias{Krumholz10} models at 1 and 10~\Msol.  (Note that
these models span all masses, and do not require normalization
for comparison with observations as the 
\citetalias{Masch10} measurements do.) 

The most noticeable feature of the \citetalias{Krumholz10} models
is that the fraction of stars expected at 10~\Msol\ and above
is relatively high for both low and high density environments.
All of the \citetalias{Krumholz10} models predict more stars at
10~\Msol\ and above than even the \citetalias{Masch10} distribution
for groups; the \citetalias{Krumholz10} range lies well
above the observed value for the four nearby star-forming regions,
and is barely consistent with the high surface density / grouped stellar
population (only) in the ONC.
The predicted fraction of massive stars in the \citetalias{Krumholz10}
models could be decreased by assuming a 
smaller maximum core mass, or, in the low density case, 
by assuming either a smaller fraction of the mass forms the
main source, or that the core fragments into a 
larger number of pieces. 

\subsection{Comparison to Field Star MF}
\begin{figure}[htb]
\plotone{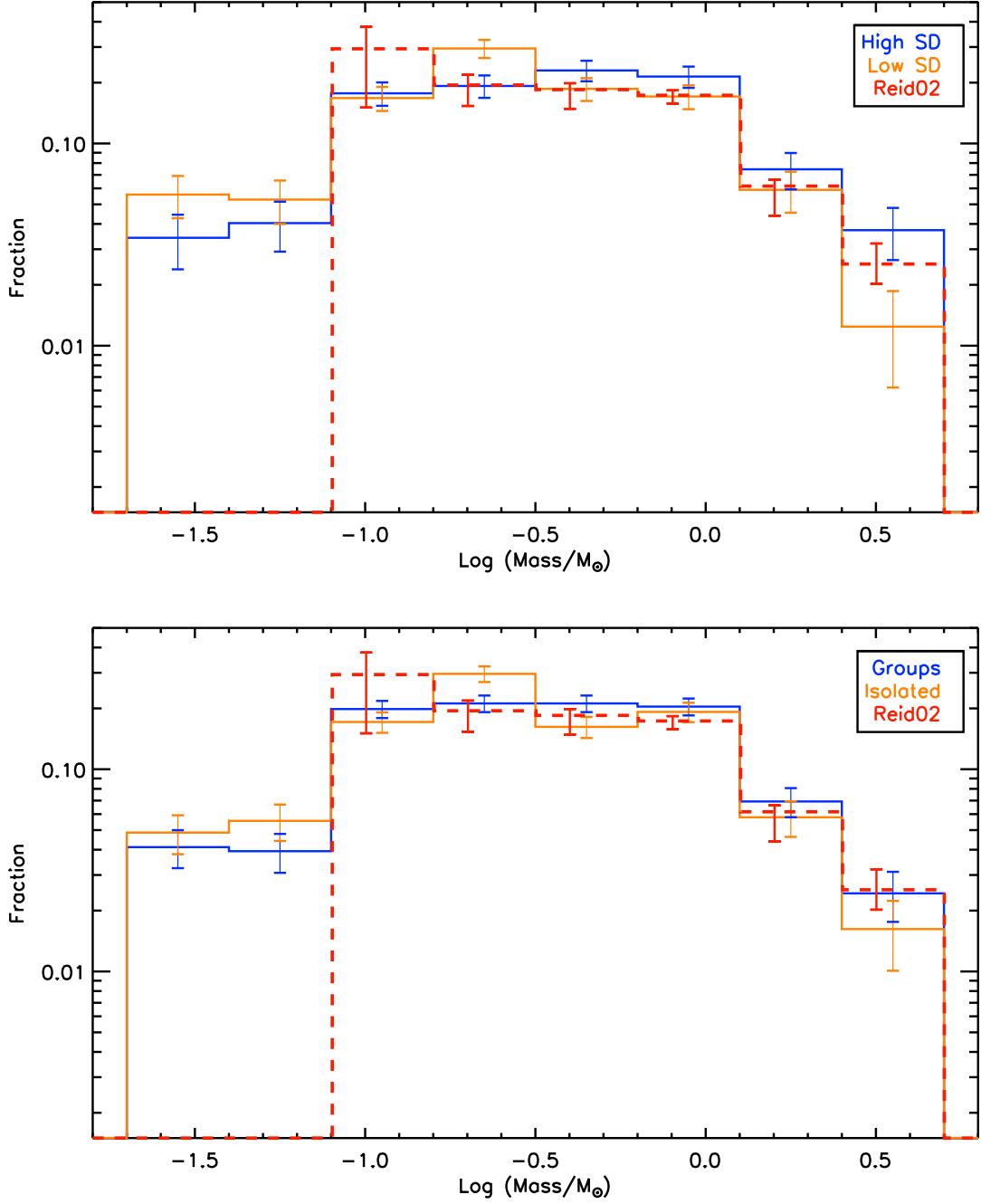}
\caption{The differential mass distribution for all four nearby
	star-forming regions.  The top panel shows the high and
	low stellar surface density stars, while the bottom
	panel shows the grouped and isolated stars.  Error
	bars indicate Poisson errors.  Also shown is the field star
	mass function measured by \citet{Reid02}, corrected
	for stellar evolution (red line).  The errors estimated by
	\citet{Reid02} are shown offset from the bin centre for
	clarity.}
\label{fig_combo_diffl_mass}
\end{figure}

Finally, we compare to observed mass functions.
Mass functions are often presented in differential form,
i.e., number per mass bin, so we follow that convention here.
Figure~\ref{fig_combo_diffl_mass} shows the differential
mass function for high and low surface density stars (top panel)
and grouped and isolated stars (bottom panel).
The general tendency is for the high surface density 
stars to be more numerous than the low surface density  
stars at the upper mass end, and the reverse at the low mass
end.  Similarly, the grouped stars tend to be more numerous than
the isolated stars at high mass, and less numerous at low
mass.  This common behaviour is more prominent for stars at
high and low surface density than for stars in groups and
isolation, as was seen earlier.

We note that inaccuracies due to our assumption of
a constant stellar age are more apparent in differential than
cumulative form.  In 
Figure~\ref{fig_combo_cuml_mass}, assuming an age of 1~Myr or 2~Myr
when estimating the masses above 1~\Msol\ creates barely
discernable differences in cumulative mass distributions. 
In the differential mass distribution, however, 
there is a noticeable shift of objects from the highest mass bin
to the second highest mass bin when the ages are assumed to
be older.  Most of these stars which shift bins have masses
little above the boundary between the two bins with an
assumed age of 1~Myr; our
bin sizes and boundaries were chosen to match those in \citet{Luhman09}
to facilitate easier comparison between the two.

Keeping in mind that the estimated masses are still somewhat
uncertain, we compare the distribution of
masses in the four nearby star-forming regions to the
field star mass function.  The field star mass function provides an 
approximate measure of
the current, nearby isolated star-forming population.  These stars
are expected to have formed over a range of times, and therefore
at the present day will be somewhat depleted of their higher-mass members, 
relative to the initial mass function.
Furthermore, the field star population is expected to consist
of a combination of stars which formed in isolation, with
those that formed in small groups or clusters that have
since dispersed or ejected some of their members.
The field star mass function would thus be expected to be formed from 
a combination of our more and less clustered stellar populations.

Figure~\ref{fig_combo_diffl_mass} shows the
\citet{Reid02} field star mass function,
corrected for the effects of stellar evolution
(their Figure~14, bottom panel); this field star mass function 
spans a larger range in masses than
some more recent observations such as \citet{Covey08}.
The evolution-corrected field star mass function is
consistent, within errors, with other determinations of the IMF
\citep{Reid02}.
While the mass distribution of the stars in high surface
density environments appears to be skewed to slightly higher
masses than the \citet{Reid02} field star mass function, they are consistent
within errors.  The low surface density, grouped, and isolated
stellar mass distributions all appear to be in closer agreement
with the \citet{Reid02} field star mass function;
the largest discrepancy
is around 0.2~\Msol, where we earlier noted our mass
estimates were biased relative to those where a range of
ages are used.

\subsection{Maximum Versus Total Group Mass}
Another property of the groups examined in \citetalias{Kirk11}
was the relationship between the mass of the most massive group
member and the total group mass.
In large clusters, the mass of the maximum mass member and the total
cluster mass are correlated \citep[e.g.,][]{Weidner10}.  
At high masses, the flattening of the relationship
between maximum and total mass was interpreted as evidence 
that a maximum stellar mass exists.
At lower masses, the relationship has a constant slope which is
roughly consistent with the relationship
expected for cluster members randomly sampling the IMF,
although \citet{Weidner10} find that the scatter in the
cluster data is smaller than expected for pure random sampling.
We showed in \citetalias{Kirk11} that the relationship between
maximum and total mass continues to the 
smallest of the groups we identified.  Here, we examine whether
the trend is also present in even smaller grouping of stars, and
whether there is a minimum group size at which the correlation is seen.

\begin{figure}[htb]
\begin{tabular}{c}
\includegraphics[height=9.6cm]{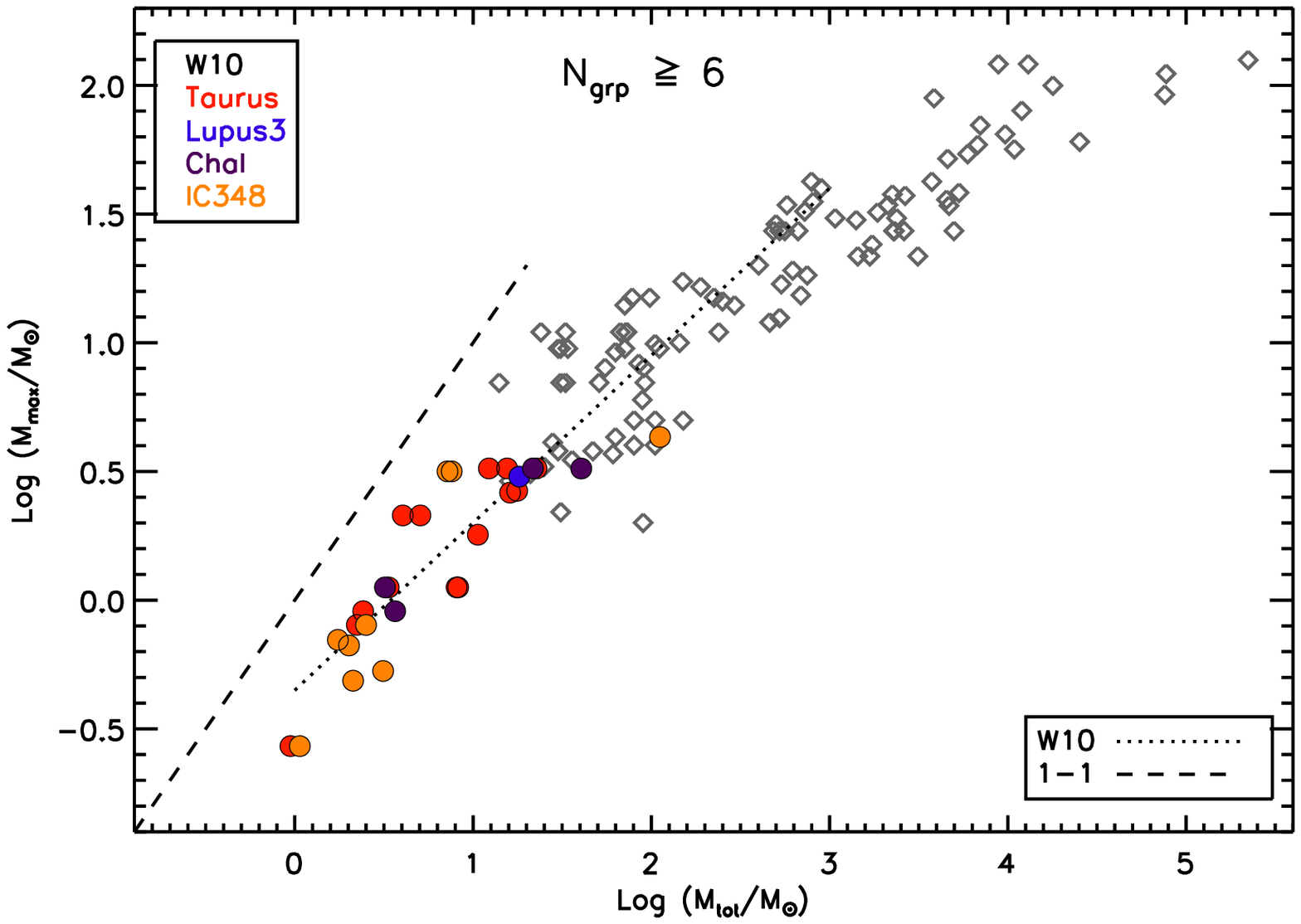}\\
\includegraphics[height=9.6cm]{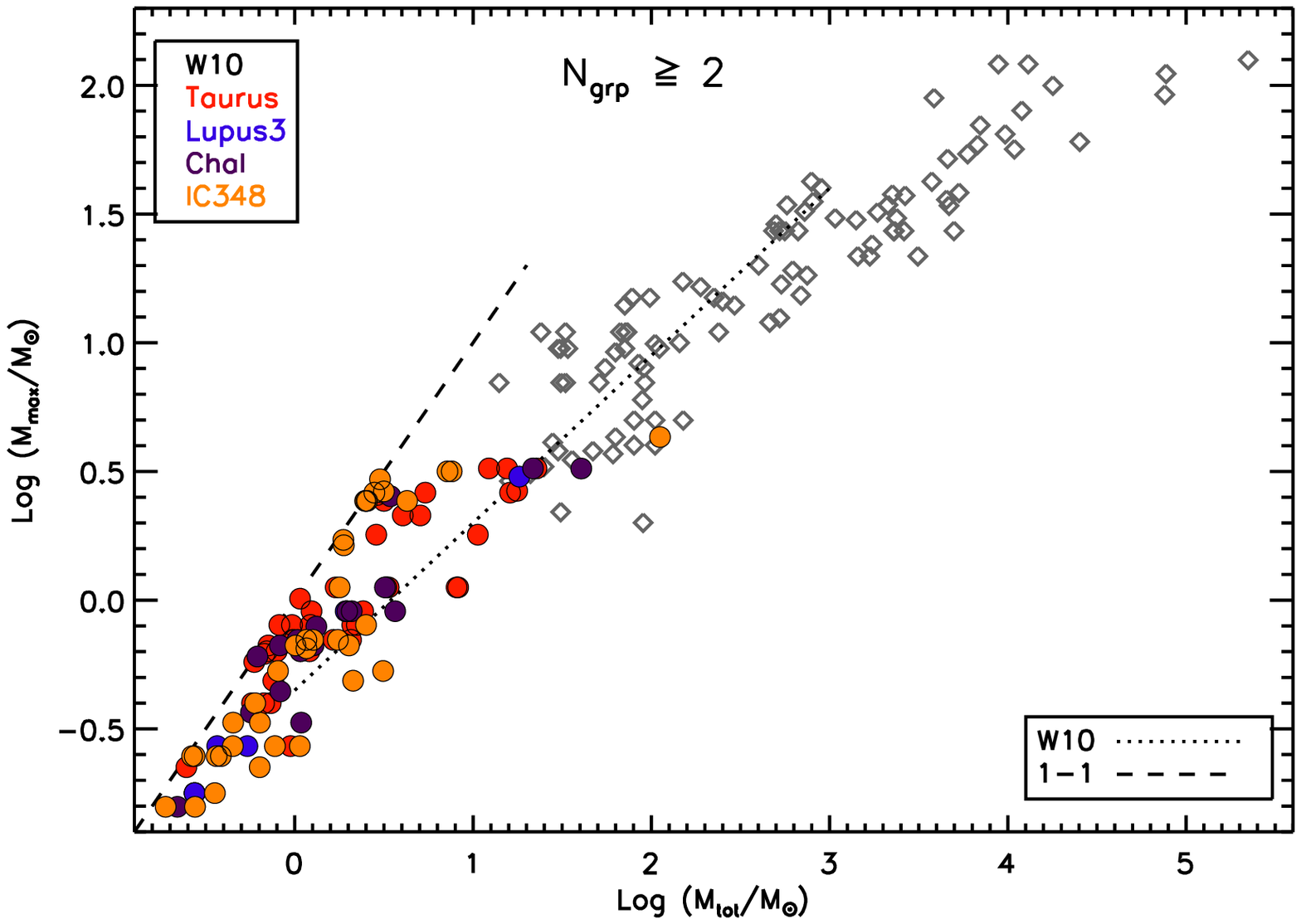}\\
\end{tabular}
\caption{The mass of the maximum mass member in each stellar grouping
	versus the total group mass.  The data and approximate low-mass
	slope in \citet{Weidner10} are shown by the grey diamonds
	and dotted line respectively.
	The coloured circles show our results for groupings with
	six or more members (top) and two or more members (bottom).
	The dashed line shows a 1-1
	relationship (the upper limit).}
\label{fig_max_vs_tot_mass}
\end{figure}

The top panel of Figure~\ref{fig_max_vs_tot_mass}
shows the relationship between maximum and total stellar mass for 
all groupings of six or more stars in Taurus, Lupus3, ChaI, and
IC348, while the
bottom panel shows the same relationship for groupings of two or
more stars.  In both panels, the grey diamonds show the cluster data
compiled by \citet{Weidner10}, while the coloured circles show the 
stellar groupings we identified with the MST.
The dotted line shows the approximate relationship fit by
\citet{Weidner10} at low cluster masses, which provides a 
reasonable description of the groups as well.
This continues to agree well with the smaller groupings, all
the way down to pairs of stars.  

While the small groupings
follow the trend remarkably well, it is worth nothing that
the range of values they can span becomes increasingly restricted
at lower group sizes.
The dashed line indicates a 1-1 relationship, the
upper limit for any group size, representing all mass residing in the
most massive group member.  For every number of group members,
there is also a lower limit, with a slope of the inverse of the
number of members; this lower limit lies parallel and below the
1-1 line on the plot.  
Even with this consideration of the smaller range possible for
small groupings, the correlation between the maximum and total
cluster mass is quite good, which suggests that properties
of stellar groups may extend to systems with fewer than ten
members.  This could be interpreted as evidence that
the masses of stars in the small groupings are consistent
with a random sampling of the IMF or an IMF-like distribution,
even though the total number of members is too small to 
measure this directly.
Higher accuracy mass estimates would be needed to determine
the consistency of the scatter in the data with that expected
from pure random sampling.

\section{DISCUSSION}
In clustered regions, the tendency to find more massive stars in
regions of higher surface density is linked to mass segregation,
since higher surface densities correspond to smaller cluster-centric
radii.  \citet{Hillenbrand98} observed mass segregation in the ONC
for stars with masses above 1-2~\Msol\ at all radii, and some
evidence for mass segregation at even smaller masses in the inner radii;
a more recent analysis of the ONC data with a MST-based
technique has found strong evidence of mass segregation only
down to 16~\Msol, with weaker evidence of mass segregation down to
5~\Msol \citep{Allison09}.
Similarly, mass segregation has been observed for IC348 
\citep[e.g.][]{Muench07}.  In \citetalias{Kirk11}, we analyzed
mass segregation from the perspective of the groups, and found
that the most massive group member, and in a few cases, also
the second most massive group member, tended to be centrally located,
reminiscent of mass segregation.  

In this work, we extend the analysis to include stars not in
groups, and find that the differences in mass distributions apply
not only to clusters, but also to dense and sparse groups.  This
tendency suggests a kind of ``mass enrichment'' associated with
stellar surface density, but not ``mass segregation'' due to 
location within a group.
We find that massive stars
tend to be more prevalent in higher stellar surface density
environments and within stellar groupings, well below the
surface densities typically associated with clusters.  Perhaps
the most surprising result is that this trend is not only strong in
IC348 and the ONC where it had been previously observed, but
is also seen in the relatively dispersed stellar population of Taurus.
Given the tendency for more massive sources to be found in 
more clustered environments in both the clustered IC348 and
the dispersed Taurus regions, it is puzzling that a similar
trend was not seen in ChaI, whose environment appears to be
intermediate between these two extremes.  Two of the three
groups identified in ChaI did, however follow the general
trend found it \citetalias{Kirk11} of having a centrally-located
most massive group member. 

Stellar motion is the final piece of the puzzle not yet 
examined.  Since all of the regions are quite young, the
stars should not have had a chance yet to move a significant
distance away from their birth sites.  We consider the available
proper motion data for all of the early type stars in our sample
to verify this assumption in Appendix~\ref{app_pm}.  Within the
current observational uncertainties, there is no evidence for
stellar motion to have influenced our results. 

How do the differences in the stellar distributions in more
versus less clustered environments arise?  Are more
low mass stars preferentially born in lower density
environments, or are they former members of small groups
that were cast out at an early age?  Our data show
suggestive signs that the excess in massive
stars in more clustered environments may not be a
uniform excess, but may be clustered around several spectral
types, specifically an excess of late B types and late K types,
corresponding to masses around 3 and 1\Msol\ at 1~Myr.  
If this observation
were borne out in a larger dataset with higher statistical significance,
that would suggest that a more complicated physical process related
to formation mechanisms is responsible.  It is difficult to see
how a simple picture of ejection, for example, would have such
specific preferred mass scales.
If instead there is a roughly uniform excess of massive 
stars in more clustered environment, this could be more easily attributed 
to the formation process.  The fact that many stars form in a
group suggests that the group environment has advantageous conditions 
relative to the field environment,
such as a gas resevoir which allows for higher accretion
rates or longer accretion timescales, which would be favourable for
forming higher mass stars.

The differences seen between the masses of clustered and isolated
young stellar objects suggests that the
initial mass function (IMF) is not 
independent of the physical conditions of star formation.
Differences in the IMF may not always be so
apparent, as the primary contributions are likely to be from
the most `successful' star-forming regions, i.e., more
clustered environments.

\section{CONCLUSION}
Complementary to the analysis presented in \citetalias{Kirk11}, we
compare the more and less clustered populations of stars found in  
four nearby young star-forming regions -- Taurus, Lupus3, ChaI, and
IC348.  The clustered young stellar population in the ONC
is also used as a comparison.  We find the following results:
\begin{enumerate}

\item Stars in the most clustered of regions, IC348 and the ONC,
have statistically significant differences between the
populations associated with high and low surface density and also
with groups and isolation.  The stars in high surface density
environments or groups tend to be skewed
overall to more massive members at the 98\% or higher confidence 
level.  These statistics are driven by by differences in the
populations at the low-mass end, although differences
are seen out to the earliest spectral types as well.
The typical difference in masses between the high surface density or 
grouped stars and the low surface density or isolated stars
is $\sim$11-18\% in IC348.

\item Despite being the classic example of isolated, distributed
star formation, the stars in Taurus follow the same trends as
the clustered regions.  Stars in higher surface density environments
in Taurus tend to have larger masses than the lower surface density
environment stars at the 99\% confidence
level, and stars in groups tend
to have larger masses than those found in isolation
at the 96\% confidence level.  These statistics
are again primarily driven by differences in the populations
around spectral type M, although differences are seen at 
the earliest spectral types.  The typical difference in masses
between the high surface density or grouped stars and the low surface
density or isolated stars is $\sim$4-18\%.

\item The Lupus3 catalog is not large enough to 
allow for any strong conclusions to be made.
\item ChaI does not follow the trends seen in the
other regions, and shows no evidence for more early
type stars in grouped or higher surface density environments.
In \citetalias{Kirk11}, two of the three groups in ChaI did
follow the general trend of a centrally-located most massive
member, however.
\item The combined mass distribution from all four regions
is consistent to what is observed in the evolution-corrected
field star mass function, or, similarly, the IMF.
\item In all four regions, the relationship between the mass
of the most massive group member and the total group mass
seen in clusters and larger groups continues to arbitrarily 
small group sizes.
\item  
Stellar motion does not appear to be responsible for the ejection
of all the isolated early type stars, given the young ages
of the star-forming regions, and the large separations of some
of the early type stars from any stellar groupings.  
\end{enumerate}

The complete stellar catalogues
available for these four nearby star-forming regions allow
the properties of stars in small groups and isolation to
be quantified to an extent not previously possible.
Although theories of massive and clustered star formation tend
to focus on large, high density systems, our results suggest
that small, relatively sparse, nearby regions
offer an alternative regime in which to
test the models.  The deep, uniform completeness levels and lack of
contamination and source confusion allow
for the characterization of grouped and isolated sources at a
level not possible in more distant clusters.

\acknowledgements{The authors thank Claudia Cyganowski, 
Lynne Hillenbrand, Doug Johnstone, Charlie Lada, Kevin Luhman, Eric Mamajeck,
Tom Megeath, and Qizhou Zhang for interesting and useful discussions about
this work.  We also thank the referee for a thoughtful and thorough
report which strengthened our results.
This research has made use of the SIMBAD database, operated at CDS, 
Strasbourg, France.}

\appendix

\section{PROPER MOTION}
\label{app_pm}
The current YSO positions must be assumed to be similar to
their birth locations in order to attribute the differences in stellar
populations to the star formation process.  Significant stellar
motions are known to occur in some systems, however.
In large clusters, the ejection of even relatively massive
members is seen in numerical simulations \citep[e.g.,][]{Vine03}, 
and in observations \citep{Goodman04}.  Smaller, sparser
systems better corresponding to our stellar groups are less
well-studied.
N-body simulations of very small
groups (with N$\sim$5) including the effects of
gas accretion and drag indicate that single stars
up to one fifth the mass of the total group mass can
be ejected early in the group's evolution 
(Delgado-Donate, Clarke, \& Bate 2003; see  
Kiseleva et al 1998 for the case of no gas).
\citet{Bate03} furthermore argue that larger dynamical perturbations
occur for both single stars and binaries in larger
systems where larger-scale star-star interactions occur.
Despite the young age and low stellar surface density
in our star-forming regions, these simulations suggest that
stellar motions should be examined to determine if they
have could have effected the currently observed YSO positions.

We examine the likelihood of significant stellar motion
for the earlier type stars in our sample, both from a 
theoretical perspective of the necessary energetics 
(Appendix~\ref{app_pm}.1) and using the available
proper motion data (Appendix~\ref{app_pm}.2).
For ease of analysis, we divide the YSOs into
our standard $N > 10$ groups, and consider only the motion of the
early type stars.  The groups in general have good correspondence
to high surface density environments, so the results should be
generally applicable.  Due to their fewer numbers, the migration of 
a single early type star would have a larger impact on the resultant 
distributions than a later spectral type star; observationally, 
proper motions are only available for the brightest stars.

\subsection{Theory}
\subsubsection{Movement Out of Groups}
We first examine the likelihood that the isolated early type stars
formed within their nearest group and subsequently migrated to their current
locations.
The escape speed required to leave each group is roughly \begin{equation}
V_{esc} = \sqrt{\frac{2 G M_{tot} } {R} },
\end{equation}
where $M_{tot}$ is the current total mass of the group,
and $R$ is the separation from the group centre from which
the star would start its migration.
We consider for two cases for $R$ -- a star forming at a `typical'
group separation, or $R_{rms}$, the standard deviation of all current
group member separations, and a star forming at the group outskirts,
or $R_{max}$, the current maximum group member separation.
The values of $R$ and $V_{esc}$
for each group are given in Table~7.

\begin{figure}
\includegraphics[height=19cm]{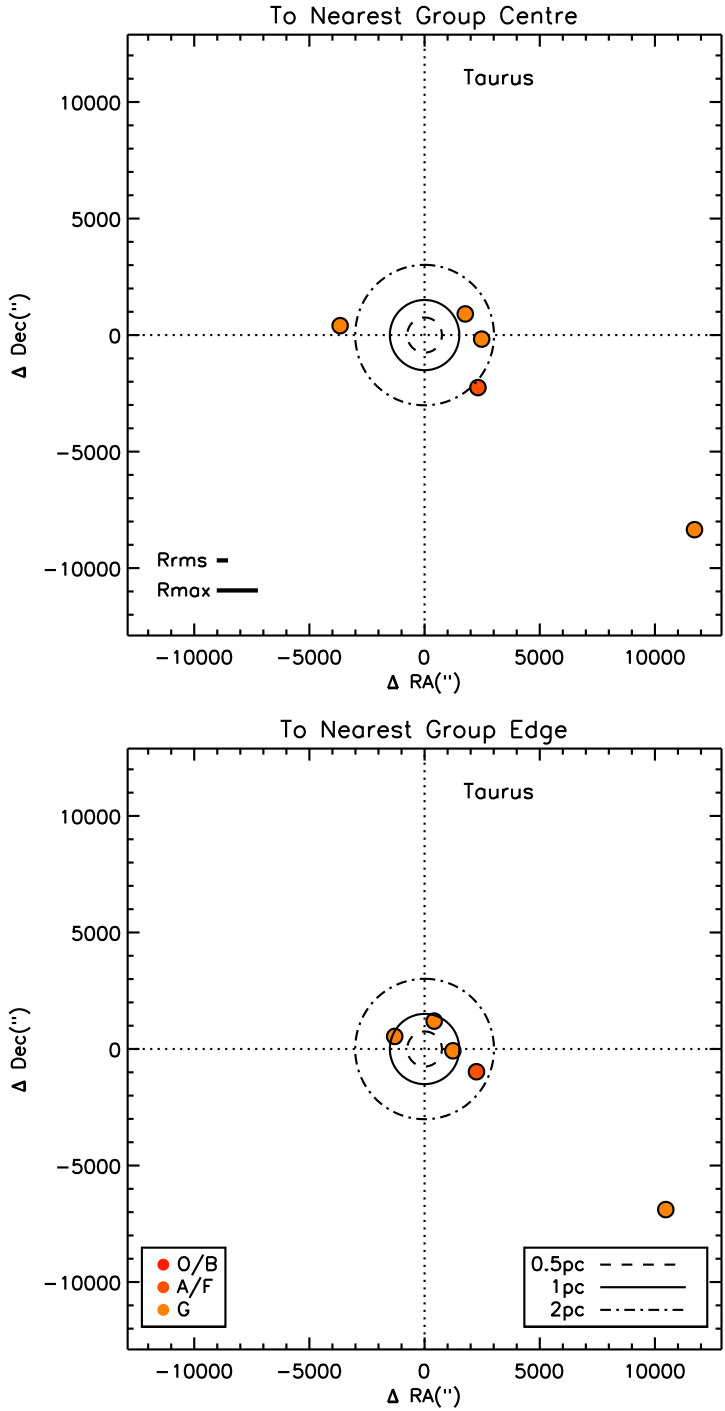}
\caption{The separation between each isolated early type star and its 
	nearest group's centre (top) and the closest group member 
	(bottom) in Taurus.  Colours
	denote different spectral types, and the circles indicate separations
	of varying amounts.  The thick horizontal lines show the mean
	Rrms and Rmax values for the groups.}
\label{fig_motion_isol_1}
\end{figure}

For each isolated early type star, we 
calculate the star's separation to both the centre
and outskirts of its nearest group, using the separation to the closest 
group member as a proxy for the latter.  For IC348, separations
only to the main cluster are considered, since it strongly
dominates the dynamics of the region; Lupus3 has no isolated
early type stars, and is not considered.  The separations are
listed in Table~8; for stars 1~Myr old,
each separation (in pc) is approximately equal to the velocity
(in km~s$^{-1}$) required to move that distance.  
The ratios between the velocity required
to reach the group centre or edge and the associated escape
velocity are also given in Table~8, along with
whether each star is more massive than
all of the current group members.
Figure~\ref{fig_motion_isol_1} shows the separations of all
isolated early type stars in Taurus from their closest
group centre (top panel) and group outskirts (bottom panel)
in right ascension and
declination, with darker shades indicating earlier spectral types.
The mean values of $R_{max}$ and $R_{rms}$ for all of the groups in
Taurus are shown in the bottom left corner of the
top panel.  The concentric annuli indicate separations of
0.5, 1, and 2~pc respectively.  The other regions
show similar behaviour, with some early type stars
well-separated from the nearest group, and others
located much closer.

Quantitatively, for a plausible dynamical `ejection' from
the group, the isolated star should be
less massive than the most massive member currently in
the group, and the velocity required to take the
star to its current location (assuming an
age of 1~Myr) should not be substantially larger than the escape 
speed for the group.
With these considerations, we see that all of the isolated
early type stars in IC348 could have moved to
their current locations, as well as few of the G stars
in Taurus and ChaI.  
Taurus-312, Taurus-352 and ChaI-209 and Cha-210,
all appear too isolated and / or massive to be potential
migration candidates.
We stress that the migration candidate stars 
undoubtedly include a significant fraction which did not
undergo migration.
Our simple analysis uses only separations on the plane of the
sky, while the 3D separation could be much larger.
Even if the 3D separation is reasonable for migration,
it does not mean that migration necessarily occured.
Further constraints using available proper motion data
are given in Appendix~\ref{app_pm}.2.

\subsubsection{Movement Into Groups}
For completeness, we also look for indications that the
early type stars currently in groups migrated there after
forming in isolation.
Table~9 gives the separation between each 
star and the group centre and edge, along with whether
it is the most massive member of its group. 
As was found in \citetalias{Kirk11}, the early type stars
tend to be found near the centre of their group,
making it unlikely that they migrated there in only
a $\sim$1~Myr timescale.  A few stars, such as Taurus-350 and Taurus-351
are the most likely candidates to
have formed either outside the group, or on the
outskirts of the group.

\subsection{Proper Motion Data}
We now use proper motion measurements in conjunction with the theoretical
considerations above to constrain the likelihood of stellar migration.

To find the proper motion of each star, we searched the recent 
all-sky proper motion catalogs of 
\citet{Zacharias10}, \citet{Roser10}, \citet{Roser08},
\citet{vanLeeuwen07}, \citet{Zacharias04}, and \citet{Kharchenko01}.
In a non-negligible number of sources, the
proper motion listed for the same source in multiple
catalogs differed by more than the errors listed.
In IC348, 
\citet{Zacharias10} list a subset of the stars as
moving with a substantially different proper motion in RA 
($\sim$25~mas~yr$^{-1}$) than is given in the other catalogs 
(most give negative values). 
In order to minimize bias from discrepant catalogs, we
use a two-step process to estimate the 
`best' proper motions for each early type star, and caution that
these values are still highly uncertain.
We first calculate the median proper motion
for each star, and then eliminate all measurements which
differ by more than 15~mas/yr from this value (the
quoted errors are typically much smaller than this).
With the remaining measurements, we calculate the weighted
mean (weighting by the error of each measure), which we will
refer to as the `best' value, and error.  
We also calculate the range spanned by all of the measures
remaining after the first step; this range of `good' proper
motion measures provides a more realistic
sense of the possible range in values than the formal error.
Tables~10 and 11 list 
the `best' proper motion, associated error, and range of
possible values for the early type stars found in isolation
and groups respectively.  Note that the proper motion
measurements are all given relative to each star's 
nearest or associated group.

The proper motion of each early type star needs to be considered
in relationship to the nearest or associated group.
Table~7 lists the proper motion we adopt for
each group.  In Taurus, we use the values given in \citet{Luhman10}; 
the \citet{Luhman10} proper groups generally match well with the
groups we identify, although both our Groups 2 and 3 match
the same \citet{Luhman10} group.  For Lupus3
and ChaI, the group proper motions have been calculated
in the same manner as for Taurus but are not yet published
(E. Mamajeck, priv. comm., Mar 9, 2011).  No statistically
significant differences were seen for the stars in the northern
and southern parts of ChaI, so the value for the entire population
is used.  A similar proper motion measurement has not yet been
made for IC348, so we estimate this value using the weighted mean
of the individual `best' proper motions we calculate for
group members; the error given is the standard error.

\subsubsection{Movement Out of Groups}
We estimate where each isolated star would have been 1~Myr ago
relative to its nearest group, using the range of good proper
motion values.
The range of possible positions is often
quite large; we list the smallest separation from the group
in this range, $Sep_{1Myr}$, in Table~10 to indicate whether
the star could possibly have originated from the group, given
the available proper motion data.
In IC348, the results are particularly
uncertain, given both the larger scatter between measurements
discussed earlier, as well as the small angular scale spanned by
the stars (small relative motions have a larger impact
on the final relative position).  Given these caveats, we
see that about half of the isolated early type stars have 
proper motions consistent with having originated in a group,
while the other half do not.  
Including the theoretical considerations in Appendix~\ref{app_pm}.1,
the stars which appear to be the most promising candidates for migration
are Taurus-97, ChaI-91, and all of the stars
in IC348 barring IC348-7 and IC348-11.

\subsubsection{Movement Into Groups}
We similarly estimate where each grouped early type star
would have been 1~Myr ago relative to its current group;
these values are given in Table~11.
Note that since $Sep_{1Myr}$ measures the predicted
minimum separation to the group {\it centre} 1~Myr ago, small,
but non-zero values are consistent with the star being a group
member in the past.  With this consideration, the only stars
whose proper motions suggest they formed outside the group and migrated
in are Taurus-58, Taurus-351, Lupus3-22, Lupus-23, ChaI-154, 
and IC348-18.

Upon closer consideration of the data, all of these sources appear to be 
consistent with originating in their current groups.
Proper motion measurements consistent with group membership
are found for both Taurus-58 and Taurus-351 in catalogs which
were excluded from the `good' values (\citet{Roser08} and 
\citet{Kharchenko01} respectively); Taurus-351 is also a binary star, 
which may contribute to the discrepancy between catalogs.
The range of good proper motion measures for both of the Lupus3 stars, 
ChaI-154, and IC348-18, are {\it nearly} compatible with the group's
proper motion; given the scatter between catalog values, it seems
likely that the errors are somewhat underestimated.

\subsection{Summary}
Significant stellar motion appears unlikely for
most of the early type stars in our sample.
A precise determination is difficult, given the large uncertainties
in the proper motion measurements, the unknown line of sight distances, and
precise ages of the stars.
Isolated early type stars in IC348
have the highest likelihood of having originated in a group, 
since this region is the most compact and tends to have the
most uncertain proper motion measurements.  In Taurus and
ChaI, a few G-type stars currently found in isolation are the
most likely candidates for having formed in a group.  None of
the stars currently in groups show compelling evidence for
having formed in isolation and migrating inward.  
The sources which may belong to a different category than
when they formed compose a small number of the total number
of early type sources, and have primarily slightly later
spectral types, where the total number of sources is larger.  Our
statistical analyses are therefore expected to remain similar
even with proper motion considerations.

Notably, the isolated F0 and B6.5 stars in ChaI appear 
unlikely to have originated in a group, suggesting that,
while rare, some
massive young stars may indeed form in isolation.

\section{EFFECT OF GROUP DEFINITION}
\label{app_mst}
In Section~\ref{sec_grp_isol}, stars were classified as
belonging to a group or being in isolation based on
the stellar groups defined in \citetalias{Kirk11}, i.e.,
stellar groups needed more than ten members, each separated
from their nearest neighbour by less than the critical
length $L_{crit}$.  We examine the effect of adopting a 
different minimum number of members to define `grouped' and
`isolated' YSOs in Appendix~\ref{app_mst}.1 and \ref{app_mst}.2,
and finally, using a different $L_{crit}$ in \ref{app_mst}.3.

\subsection{Smaller Groups}
We re-run our comparisons of grouped and 
isolated YSO spectral types using minimum group sizes of 
$N \geq $10, 8, 6, 4, and 2.  We tend to find similar
statistical results for the smaller minimum group
sizes, with the variation increasing at the smallest
group sizes.  Table~6 gives the KS2 and MW 
probabilties for a minimum group size of $N \geq 2$. 
The grouped and isolated stars in ONC are clearly still distinct 
with this drastic change in group defintion.  In Lupus3 and
ChaI, as before, the probabilities do not allow us to make 
definitive conclusions.  In IC348, the grouped and isolated 
populations appear to be much more similar with $N \geq 2$ than $N > 10$,
which appears to be due to the small number of stars which
are classified as isolated (the grouped stars are nearly seven
time more numerous than the isolated stars in this instance).
For $N \geq 6$, the KS2 and MW statistics are in better
agreement with the standard $N > 10$ case.
In Taurus, the populations of grouped and 
isolated stars actually appear to be {\it more} distinct 
for $N \geq 2$ than with $N > 10$.  
This is in agreement with the finding in Section~3 that small
groupings of stars in Taurus are distinct from
the truly isolated stars.

\subsection{Larger Groups}
Conversely, we examine the effect of {\it increasing} the minimum
group size.  Since there are few groups of any given size,
we examine minimum group sizes of $N \geq 15, 20, 25, 40,$ and 55,
chosen to match the discrete group size values in our sample.
These values also encompass group sizes in which the number of
grouped and isolated sources are roughly equal, minimizing 
the statistical uncertainty in comparisons.
The grouped and isolated source counts are closest for
$10 < N \le 16$ (Taurus), $12 < N < 36$ (ChaI),
$11 < N < 186$ (IC348), $37 < N < 410$ (ONC-c),
while in ONC-f, $N < 10$ would be required.  Lupus3 has only
one group with $N > 10$, so cannot be tested for an increase
in the minimum group size. 

Table~6 shows the KS2 and MW probabilities
for $N \geq 15$.
As the minimum group size, $N$ is increased, the KS2 probabilities 
that the `grouped' and `isolated' spectral types are
drawn from the same parent distribution tend to increase slowly
with small increases in $N$, and then rapidly increase once a significant
fraction of the original $N > 10$ grouped stars become classified as 
`isolated'.  This transition tends to occur at or slightly below the value of
$N$ which makes the number of sources classified as grouped or
isolated roughly equal.  From this, we conclude that a slight
increase of the minimum group size does not have a significant
impact on our results.  Increasing the minimum group size significantly 
($\sim$20 or higher, depending on the region), however, serves to wash out the
differences seen between grouped and isolated sources.

\subsection{Different $L_{crit}$}
$L_{crit}$, the maximum separation between 
group members, is the second variable which controls whether sources
are classified as belonging to a group or not.  In \citetalias{Kirk11},
we found an uncertainty of $\leq$ 10\% in $L_{crit}$, using
the definition in \citet{Gutermuth09} for $L_{crit}$.
Appendix~D.1 of \citetalias{Kirk11} examined the effect
of a varying $L_{crit}$ on group membership, and found typically
only a handful of late type stars would be included or excluded
from all of the groups in a region with a different $L_{crit}$.
Changes much larger than the measurement error (at least 40-60\%,
but usually at least double the measured value) are required
to cause significant changes to the classification of sources.
We therefore conclude that the choice of $L_{crit}$ does
not have a strong influence on the resulting spectral type
distributions or the statistical measures that we present.

\begin{deluxetable}{ccccccc|cccccc}
\tabletypesize{\footnotesize}
\tablecolumns{13}
\tablewidth{6.5in}
\tablecaption{Statistics for YSOs separated by surface density \label{tab_surf} }
\tablehead{ 
\colhead{ } &
\multicolumn{6}{c}{n = 4} &
\multicolumn{6}{c}{n = 9}\\
\hline
\colhead{Region} &
\colhead{KS2\tablenotemark{a}} &
\colhead{MW\tablenotemark{b}} &
\multicolumn{2}{c}{E$_{early}$\tablenotemark{c}} &
\multicolumn{2}{c}{Median $\Sigma$\tablenotemark{d}} &
\colhead{KS2\tablenotemark{a}} &
\colhead{MW\tablenotemark{b}} &
\multicolumn{2}{c}{E$_{early}$\tablenotemark{c}} &
\multicolumn{2}{c}{Median $\Sigma$\tablenotemark{d}} \\
\colhead{ } &
\colhead{ (\%) } &
\colhead{ (\%) } &
\colhead{ max $\sigma$ } &
\colhead{ cutoff } &
\colhead{low} &
\colhead{up} &
\colhead{ (\%) } &
\colhead{ (\%) } &
\colhead{ max $\sigma$ } &
\colhead{ cutoff } &
\colhead{low} &
\colhead{up}  
}
\startdata
Taurus  & 5   & 99 & 1.4 & G0,A0 & 0.65 & 44.0 & 54  & 92 & 1.4 & G0,A0 & 0.61 & 15.8 \\
Lupus3  & 96  & 75 & 1.4 & G9-G0 & 1.80 & 154  & 96  & 62 & 1.4 & G9-G0 & 2.56 & 125 \\
ChaI	& 25  & 32 & 0.7 & A5,A0 & 4.95 & 140  & 25  & 25 & 1.5 & G9 & 4.97 & 138 \\
IC348	& 3   & 99 & 3.5 & G0  & 48.3 & 667  & 0.3 & 98 & 1.8 & G0 & 48.2 & 491 \\
ONC1-c\tablenotemark{e}	& 0.1 & 99.97 & 4.9 & B5 & 28.0 & 379 & 0.005 & 99.995 & 3.0 & B5 & 27.7 & 276 \\
ONC1-f\tablenotemark{e}	& 0.8 & 98 & 7.1 & B5 & 38.7 & 977  & 5  & 90 & 7.8 & B5 & 37.4 & 910 \\
Combined\tablenotemark{f} & 0.6 & 99.8 & 3.5 & G0 & --- & --- & 5 & 97 & 2.4 & G0 & --- & --- \\
T+L+I\tablenotemark{g} & 0.2 & 99.98 & 4.5 & G0 & --- & --- & 1 & 99 & 2.4 & G0 & --- & --- \\
Combined global\tablenotemark{h} & 68 & 55 & 4.1 & G0 & 2.72 & 285 & 0.08 & 0.1 & 3.5 & G0 & 2.24 & 234 \\
T+L+I global\tablenotemark{i} & 28 & 57 & 5.3 & G0 & 2.24 & 345 & 6 & 29 & 4.1 & G0 & 2.02 & 292\\
L+C+I global\tablenotemark{j} & 9 & 98 & 6.5 & G0 & 10.9 & 425 & 22 & 91 & 6.0 & G0 & 11.3 & 339\\
\enddata
\tablenotetext{a}{The two-sample Kolmogorov-Smirnov probability that the 
	spectral types of stars in high and low surface density environments 
	are drawn from the same parent distribution.}
\tablenotetext{b}{The Mann-Whitney probability that the spectral types of
	stars in high surface density environments tend to be earlier
	than the spectral types of stars in low surface density environments.}
\tablenotetext{c}{The maximum (sigma) excess of early type stars in high versus
	low surface density environments, and the early spectral type 
	cutoff at which this occurs.}
\tablenotetext{d}{The median surface density, in $pc^{-2}$, of YSOs in
	each of the two surface density categories.}
\tablenotetext{e}{The ONC catalog using a conservative
	cut (-c) and the full catalog with no cuts (-f).}
\tablenotetext{f}{Statatistics for the combined Taurus, Lupus3, ChaI, and
	IC348 catalogs.  See Section~3.4 for details.}
\tablenotetext{g}{Statistics for the combined Taurus, Lupus3, and IC348 
	catalogs.  See Section~3.4 for details.}
\tablenotetext{h}{Statistics for Taurus, Lupus3, ChaI, and IC348, separated
	by the global surface densities.  See Section~3.4 for details.}
\tablenotetext{i}{Statistics for Taurus, Lupus3, and IC348, separated by
	the global surface densities.  See Section~3.4 for details.}
\tablenotetext{j}{Statistics for Lupus3, ChaI, and IC348, separated by
	the global surface densities.  See Section~3.4 for details.}

\end{deluxetable}

\begin{deluxetable}{cccc|ccc}
\tabletypesize{\normalsize}
\tablecolumns{7}
\tablewidth{6.5in}
\tablecaption{Effect of Spectral Type Uncertainty on Surface Density Statistics\label{tab_surf_uncert} }
\tablehead{ 
\multicolumn{7}{c}{KS2 Statistic} \\
\colhead{ } &
\multicolumn{3}{c}{n = 4} &
\multicolumn{3}{c}{n = 9}\\
\hline
\colhead{Region} &
\colhead{median \tablenotemark{a}} &
\multicolumn{2}{c}{90th percentile \tablenotemark{a}} &
\colhead{median \tablenotemark{a}} &
\multicolumn{2}{c}{90th percentile \tablenotemark{a}} \\
\colhead{ } &
\colhead{ (\%) } &
\colhead{ (\%) } &
\colhead{ (\%) } &
\colhead{ (\%) } &
\colhead{ (\%) } &
\colhead{ (\%) } 
}
\startdata
Taurus  & 8  & 4   & 11 & 54 & 34 & 76 \\
Lupus3  & 96 & 46  & 96 & 96 & 74 & 96 \\
ChaI	& 60 & 25  & 96 & 60 & 25 & 96 \\
IC348	& 3  & 0.8 &  8 & 0.5& 0.1 & 2 \\
ONC1-c\tablenotemark{b}	& 0.05 & 0.02 & 0.1 & 0.003 & 0.001 & 0.007\\
ONC1-f\tablenotemark{b}	& 0.6  & 0.3  & 1   & 3 & 2   & 6 \\
Combined\tablenotemark{c} & 0.7 & 0.2 & 2   & 6 & 2   & 14 \\
T+L+I\tablenotemark{d} &   0.2 & 0.05 & 0.7 & 1 & 0.5 & 4 \\
\cutinhead{MW Statistic}
Taurus  & 99 & 99 & 99.6 & 92 & 89 & 94 \\
Lupus3  & 75 & 60 & 96   & 62 & 48 & 74 \\
ChaI	& 35 & 26 & 45   & 27 & 19 & 36 \\
IC348	& 99 & 98 & 99.5 & 98 & 96 & 99 \\
ONC1-c\tablenotemark{b}	& 99.96 & 99.9 & 99.98 & 99.99 & 99.98 & 99.997\\
ONC1-f\tablenotemark{b}	& 97    & 95   & 98    & 88    & 83    & 92 \\
Combined\tablenotemark{c} & 99.8 & 99.6 & 99.9 & 98    & 96    & 99 \\
T+L+I\tablenotemark{d} &   99.97 & 99.9 & 99.99  & 99  & 99    & 99.7 \\
\enddata
\tablenotetext{a}{The median and the range that 90\% of the values lie within
	for the KS2 or MW statistic found with the random addition of an 
	error of half a spectral subtype.}
\tablenotetext{b}{The ONC catalog using a conservative
        cut (-c) and the full catalog with no cuts (-f).}
\tablenotetext{c}{Statatistics for the combined Taurus, Lupus3, ChaI, and
        IC348 catalogs.}
\tablenotetext{d}{Statistics for the combined Taurus, Lupus3, and IC348 
        catalogs.}
\end{deluxetable}

\begin{deluxetable}{ccccc}
\tabletypesize{\normalsize}
\tablecolumns{5}
\tablewidth{6.5in}
\tablecaption{Statistics for Grouped and Isolated YSOs \label{tab_stats} }
\tablehead{ 
\colhead{Region} &
\colhead{KS2\tablenotemark{a}} &
\colhead{MW\tablenotemark{b}} &
\multicolumn{2}{c}{$E_{early}$\tablenotemark{c}}\\
\colhead{ } &
\colhead{ (\%) } &
\colhead{ (\%) } &
\colhead{ max $\sigma$} &
\colhead{ cutoff}
}
\startdata
Taurus	& 32	& 96 & 1.4 & G0 \\
Lupus3	& 95	& 38 & 1.4 & G9-G0 \\
ChaI	& 11	& 12 & -0.4& G0 \\
IC348	& 0.8	& 98 & 1.2 & G9 \\
ONC1-c\tablenotemark{d}	& 0.02	& 99 & 1.2 & B5 \\
ONC1-f\tablenotemark{d} & 6	& 86 & 2.0 & B5 \\
Combined\tablenotemark{e} & 5 & 94 & 1.5 & G9 \\
T+L+I\tablenotemark{f} & 1 & 99 & 1.8 & G0 \\
\enddata
\tablenotetext{a}{The two-sample Kolmogorov-Smirnov probability that the 
	spectral types of grouped and isolated stars are drawn from the same 
	parent distribution.}
\tablenotetext{b}{The Mann-Whitney probability that the spectral types
	stars in groups tend to be earlier than the spectral types of
	stars in isolation.}
\tablenotetext{c}{The maximum (sigma) excess of early type stars in
	groups veruss isolation, and the early spectral type cutoff at 
	which this occurs.}
\tablenotetext{d}{A comparison to the ONC catalog, using a conservative
	cut (-c) and the full catalog with no cuts (-f).}
\tablenotetext{e}{Statistics for the combined Taurus, Lupus3, ChaI, and
	IC348 catalogs.}
\tablenotetext{f}{Statistics for the combined Taurus, Lupus3, and IC348
	catalogs.}
\end{deluxetable}

\begin{deluxetable}{cccc|ccc}
\tabletypesize{\normalsize}
\tablecolumns{7}
\tablewidth{6.5in}
\tablecaption{Effect of Spectral Type Uncertainty on MST Group Statistics\label{tab_grouped_uncert} }
\tablehead{ 
\colhead{ } &
\multicolumn{3}{c}{KS2 Statistic} &
\multicolumn{3}{c}{MW Statistic}\\
\hline
\colhead{Region} &
\colhead{median \tablenotemark{a}} &
\multicolumn{2}{c}{90th percentile \tablenotemark{a}} &
\colhead{median \tablenotemark{a}} &
\multicolumn{2}{c}{90th percentile \tablenotemark{a}} \\
\colhead{ } &
\colhead{ (\%) } &
\colhead{ (\%) } &
\colhead{ (\%) } &
\colhead{ (\%) } &
\colhead{ (\%) } &
\colhead{ (\%) } 
}
\startdata
Taurus  & 28   & 14  & 47   & 96 & 94 & 97 \\
Lupus3  & 89   & 48  & 99.7 & 38 & 26 & 53 \\
ChaI	& 28   & 9   & 63   & 15 & 9  & 22 \\
IC348	& 0.2  & 0.04& 0.8  & 98 & 96 & 99 \\
ONC1-c\tablenotemark{b}	& 0.2 & 0.04 & 0.9 & 99 & 99 & 99.6\\
ONC1-f\tablenotemark{b}	& 5   & 3    & 9   & 85 & 79 & 90 \\
Combined\tablenotemark{c} & 8 & 4    & 19  & 95 & 92 & 97 \\
T+L+I\tablenotemark{d} &    2 & 0.6  & 5   & 99 & 99 & 99.7 \\
\enddata
\tablenotetext{a}{The median and the range that 90\% of the values lie within
	for the KS2 or MW statistic found with the random addition of an 
	error of half a spectral subtype.}
\tablenotetext{b}{The ONC catalog using a conservative
        cut (-c) and the full catalog with no cuts (-f).}
\tablenotetext{c}{Statatistics for the combined Taurus, Lupus3, ChaI, and
        IC348 catalogs.}
\tablenotetext{d}{Statistics for the combined Taurus, Lupus3, and IC348 
        catalogs.}
\end{deluxetable}

\begin{deluxetable}{cccc}
\tabletypesize{\normalsize}
\tablecolumns{4}
\tablewidth{6.5in}
\tablecaption{Statistics for Individual Groups in Taurus and ChaI \label{tab_stats_indiv} }
\tablehead{ 
\colhead{Region} &
\colhead{Group} &
\colhead{KS2\tablenotemark{a}} &
\colhead{MW\tablenotemark{b}}\\
\colhead{ } &
\colhead{ } &
\colhead{ (\%) } &
\colhead{ (\%) } 
}
\startdata
Taurus	& 1 & 60 & 64 \\
Taurus	& 2 & 88 & 67 \\
Taurus	& 3 & 71 & 48 \\
Taurus	& 4 &  3 & 99.8 \\
Taurus	& 5 &  8 & 89 \\
Taurus	& 6 & 99 & 67 \\
Taurus	& 7 & 73 & 83 \\
Taurus	& 8 & 54 & 61 \\
ChaI	& 1 & 42 & 10 \\
ChaI	& 2 & 12 & 5 \\
ChaI	& 3 & 46 & 72 \\
\enddata
\tablenotetext{a}{The two-sample Kolmogorov-Smirnov probability that the 
	spectral types of grouped and isolated stars are drawn from the same 
	parent distribution.}
\tablenotetext{b}{The Mann-Whitney probability that the spectral types
	stars in groups tend to be earlier than the spectral types of
	stars in isolation.}
\end{deluxetable}

\begin{deluxetable}{ccc|cc}
\tabletypesize{\normalsize}
\tablecolumns{5}
\tablewidth{6.5in}
\tablecaption{Statistics for Groupings and Isolated YSOs with $N \geq$ 2
	and $N \geq 15$ 
	\label{tab_stats2} }
\tablehead{ 
\colhead{ } &
\multicolumn{2}{c}{$N \ge 2$} &
\multicolumn{2}{c}{$N \ge 15$} \\
\colhead{Region} &
\colhead{KS2\tablenotemark{a}} &
\colhead{MW\tablenotemark{b}}&
\colhead{KS2\tablenotemark{a}} &
\colhead{MW\tablenotemark{b}}\\
\colhead{ } &
\colhead{ (\%) } &
\colhead{ (\%) } & 
\colhead{ (\%) } &
\colhead{ (\%) } 
}
\startdata
Taurus	& 7	& 95 & 42  & 90 \\
Lupus3	& 37	& 13 & 95  & 38 \\
ChaI	& 19	& 68 & 23  & 26 \\
IC348	& 17	& 63 & 0.3 & 98 \\
ONC1-c\tablenotemark{c}	& 0.8	& 91 & 0.8 & 98 \\
ONC1-f\tablenotemark{c} & 0.02	& 81 & 0.4 & 98 \\
\enddata
\tablenotetext{a}{The two-sample Kolmogorov-Smirnov probability that the 
	spectral types of stars in any size of grouping and very isolated stars 
	are drawn from the same parent distribution.}
\tablenotetext{c}{The Mann-Whitney probability that the spectral types of
	stars in any size of grouping tend to be earlier than the spectral
	types of very isolated stars.}
\tablenotetext{c}{A comparison to the ONC catalog, using a conservative
	cut (-c) and the full catalog with no cuts (-f).}
\end{deluxetable}

\begin{deluxetable}{cccccccccc}
\tabletypesize{\footnotesize}
\tablecolumns{10}
\tablewidth{6.5in}
\tablecaption{Group Properties \label{tab_grps}}
\tablehead{
\colhead{Region} &
\colhead{Grp} &
\colhead{Rrms\tablenotemark{a}} &
\colhead{Rmax\tablenotemark{a}} &
\colhead{Vesc(Rrms)\tablenotemark{b}} &
\colhead{Vesc(Rmax)\tablenotemark{b}} &
\colhead{$\mu_{\alpha}$\tablenotemark{c}} &
\colhead{$\mu_{\delta}$\tablenotemark{c}} &
\colhead{$\Delta \mu$\tablenotemark{c}} &
\colhead{Refs \tablenotemark{d}} \\
\colhead{ } &
\colhead{\#} &
\colhead{(pc)} &
\colhead{(pc)} &
\colhead{(km~s$^{-1}$)} &
\colhead{(km~s$^{-1}$)} &
\colhead{(mas/yr)} &
\colhead{(mas/yr)} &
\colhead{(mas/yr)} &
\colhead{ }
}
\startdata
  Taurus & 1 &  0.24 &  0.82 &  0.62 &  0.33 & 6.9   & -22.3 & 1   & L10 \\
  Taurus & 2 &  0.29 &  1.13 &  0.68 &  0.34 & 6.0   & -26.8 & 1   & L10 \\
  Taurus & 3 &  0.40 &  1.31 &  0.42 &  0.23 & 6.0   & -26.8 & 1   & L10 \\
  Taurus & 4 &  0.35 &  1.38 &  0.75 &  0.38 & 10.0  & -17.6 & 1   & L10 \\
  Taurus & 5 &  0.23 &  0.69 &  0.56 &  0.32 & 5.5   & -21.9 & 1   & L10 \\
  Taurus & 6 &  0.42 &  1.64 &  0.60 &  0.30 & 6.7   & -17.7 & 2   & L10 \\
  Taurus & 7 &  0.48 &  1.85 &  0.54 &  0.27 & 4.5   & -21.3 & 2   & L10 \\
  Taurus & 8 &  0.26 &  0.88 &  0.64 &  0.35 & 3.9   & -23.4 & 1   & L10 \\
  Lupus3 & 1 &  0.24 &  1.06 &  0.80 &  0.38 & -9.4  & -21   & 1.1 & EM1 \\
    ChaI & 1 &  0.16 &  0.48 &  0.44 &  0.26 & -20.9 & 1.3   & 0.7 & EM2 \\
    ChaI & 2 &  0.21 &  1.01 &  1.27 &  0.59 & -20.9 & 1.3   & 0.7 & EM2 \\
    ChaI & 3 &  0.17 &  0.61 &  1.04 &  0.55 & -20.9 & 1.3   & 0.7 & EM2 \\
   IC348 & 1 &  0.13 &  0.58 &  2.73 &  1.29 & 5.6   & -6.6  & 0.8 & KM \\
   IC348 & 2 &  0.04 &  0.15 &  0.79 &  0.43 & 5.6   & -6.6  & 0.8 & KM \\
\enddata
\tablenotetext{a}{The rms and maximum separation of group members from
	the group centre.}
\tablenotetext{b}{Escape velocity for the group, calculated at both Rrms
	and Rmax}
\tablenotetext{c}{Proper motion measured for each group (RA, dec, and
	approximate standard error)}
\tablenotetext{d}{References for proper motion measurement.  L10: 
	\citet{Luhman10}; Groups 2 and 3 both lie within Luhman's
	proper motion Group II.  EM1: E. Mamajek, priv. comm, 
	Mar 9, 2011.  EM2: E. Mamajek, priv. comm, Mar 9, 2011; no
	statistically significant difference is seen between 
	the northern and southern stars, so the average value for
	the entire region is adopted.  KM: Calculated by the authors;
	more details in text.}
\end{deluxetable}

\begin{deluxetable}{ccccccccc}
\tabletypesize{\normalsize}
\tablecolumns{9}
\tablewidth{6.5in}
\tablecaption{Required Motions for Isolated YSOs \label{tab_isol_Vesc}}
\tablehead{
\colhead{Region} &
\colhead{Src} &
\colhead{Spec.} &
\colhead{Closest \tablenotemark{a}} &
\colhead{Separation} &
\colhead{Separation} &
\colhead{V/Vesc\tablenotemark{b}} &
\colhead{V/Vesc\tablenotemark{c}} &
\colhead{$M <$ \tablenotemark{d}}\\
\colhead{} &
\colhead{\#} &
\colhead{Type} &
\colhead{Grp} &
\colhead{Centre (pc)} &
\colhead{Edge (pc)} &
\colhead{Centre} &
\colhead{Edge} &
\colhead{$M_{max}?$}
}
\startdata
  Taurus &  97 & G5 & 2 &   1.69 &   0.84 &   2.53 &   2.49 & Y \\
  Taurus & 160 & G8 & 6 &   2.50 &   0.95 &   4.26 &   3.20 & Y \\
  Taurus & 233 & G8 & 4 &   1.35 &   0.86 &   1.83 &   2.34 & Y \\
  Taurus & 312 & G5 & 4 &   9.77 &   8.51 &  13.29 &  23.16 & Y \\
  Taurus & 352 & A2 & 8 &   2.19 &   1.67 &   3.49 &   4.92 & Y \\
    ChaI &  91 & G9 & 3 &   0.63 &   0.23 &   0.62 &   0.43 & Y \\
    ChaI & 209 & F0 & 1 &   2.56 &   2.21 &   5.95 &   8.87 & N \\
    ChaI & 210 & B6.5 & 3 &   1.38 &   0.92 &   1.35 &   1.70 & N \\
   IC348 &   3 & A0 & 1 &   0.94 &   0.44 &   0.35 &   0.35 & Y \\
   IC348 &   7 & A0 & 1 &   0.49 &   0.18 &   0.18 &   0.14 & Y \\
   IC348 &   8 & A2 & 1 &   0.48 &   0.17 &   0.18 &   0.14 & Y \\
   IC348 &  11 & G4 & 1 &   0.76 &   0.50 &   0.29 &   0.39 & Y \\
   IC348 &  17 & A4 & 1 &   0.93 &   0.44 &   0.35 &   0.35 & Y \\
   IC348 &  19 & G1 & 1 &   0.64 &   0.37 &   0.24 &   0.29 & Y \\
   IC348 &  21 & G5 & 1 &   0.87 &   0.50 &   0.32 &   0.39 & Y \\
   IC348 &  25 & A4 & 1 &   0.62 &   0.37 &   0.23 &   0.29 & Y \\
\enddata
\tablenotetext{a}{For IC348, all comparisons are
        to IC348-main; sources 3, 7, and 21 lie closer
        to IC348-North.}
\tablenotetext{b}{Ratio of the velocity required for the source to move to
	its current location from the group centre to the escape velocity
	at Rrms.}
\tablenotetext{c}{Ratio of the velocity required for the source to move to
	its current location from the nearest current group member
	to the escape velocity at the group edge.}
\tablenotetext{d}{Is the source less massive than the most massive
	group member?}
\end{deluxetable}

\begin{deluxetable}{ccccccc}
\tabletypesize{\footnotesize}
\tablecolumns{7}
\tablewidth{6.5in}
\tablecaption{Required Motions for Grouped YSOs \label{tab_grp_Vesc}}
\tablehead{
\colhead{Region} &
\colhead{Src} &
\colhead{Spec.} &
\colhead{Group\tablenotemark{a}} &
\colhead{Separation} &
\colhead{Separation} &
\colhead{$M =$}\\
\colhead{} &
\colhead{\#} &
\colhead{Type} &
\colhead{ } &
\colhead{Edge (pc)} &
\colhead{Centre (pc)} &
\colhead{$M_{max}?$\tablenotemark{b}}
}
\startdata
  Taurus &  58 & B9 & 2 & 1.01 & 0.12  &  Y \\
  Taurus & 255 & G0 & 6 & 1.27 & 0.36  &  Y \\
  Taurus & 287 & G5 & 7 & 1.44 & 0.40  &  Y \\
  Taurus & 328 & B9 & 8 & 0.68 & 0.20  &  Y \\
  Taurus & 336 & G2 & 8 & 0.62 & 0.26  &  N \\
  Taurus & 350 & B9.5 & 4 & 0.43 & 0.95  &  N \\
  Taurus & 351 & B9 & 4 & 0.43 & 0.95  &  Y \\
  Lupus3 &  21 & A7 & 1 & 1.03 & 0.03  &  N \\
  Lupus3 &  22 & A3 & 1 & 1.02 & 0.03  &  Y \\
    ChaI &  11 & G5 & 2 & 0.43 & 0.58  &  N \\
    ChaI &  53 & B9 & 3 & 0.47 & 0.13  &  Y \\
    ChaI & 116 & G2 & 2 & 0.63 & 0.37  &  N \\
    ChaI & 121 & B9.5 & 2 & 0.63 & 0.38  &  N \\
    ChaI & 154 & G8 & 2 & 0.27 & 0.74  &  N \\
    ChaI & 182 & G7 & 2 & 0.87 & 0.13  &  N \\
   IC348 &   1 & B5 & 1 & 0.51 & 0.07  &  Y \\
   IC348 &   2 & A2 & 1 & 0.48 & 0.10  &  N \\
   IC348 &   4 & F0 & 1 & 0.34 & 0.23  &  N \\
   IC348 &   5 & G8 & 1 & 0.16 & 0.41  &  N \\
   IC348 &   6 & G3 & 1 & 0.37 & 0.20  &  N \\
   IC348 &   9 & G8 & 1 & 0.47 & 0.10  &  N \\
   IC348 &  10 & F2 & 1 & 0.37 & 0.20  &  N \\
   IC348 &  12 & G0 & 1 & 0.33 & 0.24  &  N \\
   IC348 &  13 & A3 & 1 & 0.33 & 0.24  &  N \\
   IC348 &  16 & G6 & 1 & 0.54 & 0.04  &  N \\
   IC348 &  18 & A2 & 1 & 0.48 & 0.10  &  N \\
   IC348 &  28 & F0 & 1 & 0.30 & 0.27  &  N \\
   IC348 &  29 & G1 & 1 & 0.12 & 0.45  &  N \\
   IC348 &  35 & G0 & 1 & 0.32 & 0.25  &  N \\
   IC348 & 359 & B5 & 1 & 0.51 & 0.07  &  Y \\
   IC348 & 360 & G8 & 1 & 0.47 & 0.10  &  N \\
\enddata
\tablenotetext{a}{For IC348, all comparisons are
	to IC348-main; sources 3, 7, and 21 lie closer
	to IC348-North.}
\tablenotetext{b}{Is the source the most massive group member?}
\end{deluxetable}

\begin{deluxetable}{ccccccccccc}
\tabletypesize{\footnotesize}
\tablecolumns{11}
\tablewidth{6.5in}
\tablecaption{Observed Relative Motions for Isolated YSOs \label{tab_isol_pm}}
\tablehead{
\colhead{Region} &
\colhead{Src} &
\colhead{$\mu_{\alpha,B}$\tablenotemark{a}} &
\colhead{$\Delta \mu_{\alpha,B}$\tablenotemark{a}} &
\colhead{$\mu_{\delta,B}$\tablenotemark{a}} &
\colhead{$\Delta \mu_{\delta,B}$\tablenotemark{a}} &
\colhead{Min $\mu_{\alpha}$\tablenotemark{b}} &
\colhead{Max $\mu_{\alpha}$\tablenotemark{b}} &
\colhead{Min $\mu_{\delta}$\tablenotemark{b}} &
\colhead{Max $\mu_{\delta}$\tablenotemark{b}} &
\colhead{Sep$_{1Myr}$\tablenotemark{c}} 
}
\startdata
Taurus & 97 & 1.63 & 1.14 & -1.05 & 1.13 & -1.08 &  4.02 & -3.90 & 1.15 & 0\\
Taurus &160 &-5.70 & 4.60 &-27.33 & 4.58 &-32.19 &  2.12 &-44.29 &-21.18& 14.6\\
Taurus &233 &-13.26& 2.42 &  9.61 & 2.42 &-15.68 &-10.84 &  7.19 &12.03 & 9.6\\
Taurus &312 & 2.66 & 1.19 & -1.70 & 1.19 & -0.11 &  4.70 & -3.29 & 0.04 & 5.9\\
Taurus &352 & 1.25 & 1.09 & -1.53 & 1.10 & -0.97 &  3.09 & -4.29 & 0.28 & 0\\
  ChaI & 91 &-0.40 & 1.11 &  1.11 & 1.04 & -3.87 &  3.11 & -2.81 & 5.97 & 0\\
  ChaI &209 &-5.78 & 0.80 &-11.47 & 0.72 & -7.19 & -4.53 &-13.29 &-8.79 & 6.4\\
  ChaI &210 &-1.90 & 0.73 &  0.12 & 0.67 & -3.50 & -0.63 & -0.91 & 1.30 & 0.2\\
 IC348 &  3 &-2.30 & 0.93 &  0.48 & 0.68 & -3.94 &  3.94 & -2.09 & 1.41 & 0\\
 IC348 &  7 & 2.78 & 1.16 & -1.93 & 1.00 &  0.76 &  5.55 & -5.31 &-0.01 & 1.6\\
 IC348 &  8 &-0.04 & 1.16 & -0.69 & 0.98 & -1.63 &  2.38 & -5.42 & 1.23 & 0\\
 IC348 & 11 &-8.38 & 4.84 & -4.04 & 4.80 &-22.45 & -2.74 &-18.90 & 1.72 & 4.6\\
 IC348 & 17 &-2.54 & 5.45 & -2.10 & 4.41 &-30.92 &  3.76 &-10.62 &34.51 & 0\\
 IC348 & 19 &-5.34 & 2.74 & -0.66 & 4.70 & -9.10 &  2.26 & -8.32 &15.31 & 0\\
 IC348 & 21 &-5.49 & 2.27 &  2.90 & 4.04 & -8.14 &  1.16 & -7.02 &20.05 & 0\\
 IC348 & 25 &-1.45 & 3.83 & -0.16 & 4.01 & -6.76 &  4.06 & -8.82 & 9.62 & 0\\

\enddata
\tablenotetext{a}{The estimated `best' proper motion of each source relative
	to its nearest group and the associated errors, both in mas~yr$^{-1}$.}
\tablenotetext{b}{The range of `good' proper motion measurements of each
	source relative to its nearest group, all in mas~yr$^{-1}$. } 
\tablenotetext{c}{The minimum separation to the nearest group centre 1~Myr
	ago (in pc), based on the range of relative proper motion measurements.}
\end{deluxetable}

\begin{deluxetable}{ccccccccccc}
\tabletypesize{\footnotesize}
\tablecolumns{11}
\tablewidth{6.5in}
\tablecaption{Observed Relative Motions for Grouped YSOs \label{tab_grp_pm}}
\tablehead{
\colhead{Region} &
\colhead{Src} &
\colhead{$\mu_{\alpha,B}$\tablenotemark{a}} &
\colhead{$\Delta \mu_{\alpha,B}$\tablenotemark{a}} &
\colhead{$\mu_{\delta,B}$\tablenotemark{a}} &
\colhead{$\Delta \mu_{\delta,B}$\tablenotemark{a}} &
\colhead{Min $\mu_{\alpha}$\tablenotemark{b}} &
\colhead{Max $\mu_{\alpha}$\tablenotemark{b}} &
\colhead{Min $\mu_{\delta}$\tablenotemark{b}} &
\colhead{Max $\mu_{\delta}$\tablenotemark{b}} &
\colhead{Sep$_{1Myr}$\tablenotemark{c}} \\
\colhead{ } &
\colhead{\#} &
\multicolumn{2}{c}{(mas~yr$^{-1}$)} &
\multicolumn{2}{c}{(mas~yr$^{-1}$)} &
\multicolumn{2}{c}{(mas~yr$^{-1}$)} &
\multicolumn{2}{c}{(mas~yr$^{-1}$)} &
\colhead{(pc)}
}
\startdata
Taurus &  58 &-27.15 & 8.12 & -2.02 & 4.26 &-39.47 &-15.06 & -8.10 &  9.73 & 10.1\\
Taurus & 255 &  4.71 & 2.21 &  3.73 & 2.21 &  1.49 &  7.56 & -0.22 &  6.90 & 0.6\\
Taurus & 287 & -2.90 & 6.23 &  1.60 & 6.23 & -9.13 &  3.33 & -4.63 &  7.83 & 0\\
Taurus & 328 & -1.68 & 1.13 & -1.17 & 1.06 & -3.35 &  1.24 & -2.93 &  0.25 & 0.03\\
Taurus & 336 &  0.79 & 1.26 & -0.25 & 1.19 & -5.78 &  3.22 & -3.75 &  3.04 & 0\\
Taurus & 350 & -2.95 & 1.44 & -0.64 & 1.11 & -6.76 &  2.33 & -3.36 &  1.87 & 0\\
Taurus & 351 & 15.20 & 2.09 & -1.66 & 1.49 & 12.41 & 17.99 & -6.41 &  2.05 & 7.7\\
Lupus3 &  21 &  0.18 & 1.17 & -2.91 & 1.17 & -1.69 &  4.44 & -7.43 & -0.96 & 0.9\\
Lupus3 &  22 &  3.10 & 1.18 & -1.22 & 1.16 &  0.91 &  5.42 & -3.25 &  0.25 & 0.9\\
  ChaI &  11 &  2.26 & 1.01 & -1.79 & 1.78 &  0.51 &  7.48 & -4.75 &  1.18 & 0.8\\
  ChaI &  53 & -0.54 & 0.84 & -1.91 & 0.76 & -1.99 &  1.46 & -3.29 &  0.48 & 0\\
  ChaI & 116 & -3.29 & 1.36 &  3.06 & 1.28 & -5.65 &  0.38 & -1.04 &  8.05 & 0\\
  ChaI & 121 &  1.42 & 0.78 &  0.41 & 0.71 & -0.75 &  3.25 & -1.37 &  1.77 & 0\\
  ChaI & 154 & -0.69 & 1.20 &  3.52 & 1.17 & -1.89 &  0.51 &  2.35 &  4.69 & 1.4\\
  ChaI & 182 & -0.44 & 2.73 &  6.66 & 2.38 &-10.15 &  8.87 &-11.21 & 16.71 & 0\\
 IC348 &   1 & -1.03 & 1.15 & -0.15 & 0.97 & -8.04 &  0.64 & -6.77 & 16.71 & 0\\
 IC348 &   2 &  1.39 & 1.44 &  0.23 & 1.32 & -2.02 &  3.95 & -4.46 &  2.56 & 0\\
 IC348 &   4 &  2.52 & 1.21 & -0.24 & 0.87 & -4.74 &  5.30 & -2.31 &  2.56 & 0\\
 IC348 &   5 & -1.25 & 3.73 & -2.51 & 3.60 & -8.96 &  6.26 &-13.88 &  6.92 & 0\\
 IC348 &   6 & -6.58 & 3.55 & -2.21 & 4.59 &-12.77 &  1.86 & -9.72 & 22.11 & 0\\
 IC348 &   9 &  6.29 & 1.59 &  3.78 & 3.08 &-10.15 &  8.31 &-11.52 &  9.34 & 0\\
 IC348 &  10 &  0.92 & 1.10 &  3.41 & 1.12 & -1.84 &  3.80 & -0.43 &  5.09 & 0\\
 IC348 &  12 & -4.80 & 5.66 & -2.13 & 4.94 &-10.46 &  0.86 &-14.11 &  4.02 & 0\\
 IC348 &  13 & -4.80 & 5.66 & -2.13 & 4.94 &-10.46 &  0.86 &-14.11 &  4.02 & 0\\
 IC348 &  16 & -3.20 & 3.10 & -3.30 & 2.94 & -6.30 & -0.10 & -6.24 & -0.36 & 0.5\\
 IC348 &  18 &  2.91 & 1.50 & -1.12 & 1.36 &  0.93 &  4.90 & -3.00 &  0.77 & 1.4\\
 IC348 &  28 &  0.37 & 1.46 &  0.85 & 1.40 & -5.85 &  4.10 & -2.35 &  2.57 & 0\\
 IC348 &  29 & -5.20 & 6.95 & -6.00 & 6.92 &-12.15 &  1.75 &-12.92 &  0.92 & 0\\
 IC348 &  35 & -3.83 & 3.45 & -1.65 & 3.65 &-10.26 &  0.88 & -6.63 & 26.31 & 0\\
 IC348 & 359 & -1.01 & 1.10 & -0.25 & 0.92 &-12.81 &  0.64 & -3.23 & 16.71 & 0\\
 IC348 & 360 &  6.29 & 1.59 &  3.78 & 3.08 &-10.15 &  8.31 &-11.52 &  9.34 & 0\\
\enddata
\tablenotetext{a}{The estimated `best' proper motion of each source relative
	to its group and the associated errors.} 
\tablenotetext{b}{The range of `good' proper motion measurements of each
	source relative to its group.}
\tablenotetext{c}{The minimum separation to the nearest group centre 1~Myr 
	ago.}
\end{deluxetable}

\clearpage
\newpage

\newpage
	


\newpage

\end{document}